\documentclass[
aps
,prb
,twocolumn,10pt
,superscriptaddress
,longbibliography
,amsmath
,amssymb
,amsfonts
]{revtex4-2}
\usepackage{float}
\usepackage{commath}
\usepackage{graphicx}
\usepackage{verbatim}
\usepackage{color}
\usepackage[dvipsnames,svgnames,x11names,hyperref]{xcolor}

\usepackage{times}
\usepackage[compat=1.1.0]{tikz-feynman}
\usepackage[sf,tight]{subfigure}

\usepackage{siunitx}
\DeclareSIUnit\gauss{G} 

\definecolor{linkcolor}{RGB}{6,69,173} 
\definecolor{diffcolor}{RGB}{175,31,36} 

\usepackage[colorlinks=true,
            linkcolor=linkcolor,
            urlcolor=linkcolor,
            citecolor=linkcolor,
            unicode,
            pdfencoding=auto]{hyperref}

\usepackage[
]{physpack}

\newcommand{\smone}{\ensuremath{\mathord{-}1}}

\begin{document}

\title{Multimagnon dynamics and thermalization in the \texorpdfstring{$S=1$}{S=1} easy-axis ferromagnetic chain}

\author{Prakash Sharma}
\affiliation{Department of Physics, Florida State University, Tallahassee, Florida 32306, USA}
\affiliation{National High Magnetic Field Laboratory, Tallahassee, Florida 32310, USA}

\author{Kyungmin Lee}
\affiliation{Department of Physics, Florida State University, Tallahassee, Florida 32306, USA}
\affiliation{National High Magnetic Field Laboratory, Tallahassee, Florida 32310, USA}

\author{Hitesh J. Changlani}
\email{hchanglani@fsu.edu}
\affiliation{Department of Physics, Florida State University, Tallahassee, Florida 32306, USA}
\affiliation{National High Magnetic Field Laboratory, Tallahassee, Florida 32310, USA}

\begin{abstract}
Quasiparticles are physically motivated mathematical constructs for simplifying the seemingly complicated many-body description of solids.
A complete understanding of their dynamics and the nature of the effective interactions between them provides rich information on real material properties at the microscopic level.
In this work, we explore the dynamics and interactions of magnon quasiparticles in a ferromagnetic spin-1 Heisenberg chain with easy-axis onsite anisotropy,
a model relevant for the explanation of recent terahertz optics experiments on NiNb$_2$O$_6$ [P. Chauhan \textit{et al.}, Phys. Rev. Lett. \textbf{124}, 037203 (2020)],
and nonequilibrium dynamics in ultracold atomic settings [W.C. Chung \textit{et al.}, Phys. Rev. Lett. \textbf{126}, 163203 (2021)].
We build a picture for the properties of clouds of a few magnons with the help of exact diagonalization and 
density matrix renormalization group calculations supported by physically motivated Jastrow wavefunctions.
We show how the binding energy of magnons effectively reduces with their number and explain how this energy scale is of direct relevance for dynamical magnetic susceptibility measurements.
This understanding is used to make predictions for ultracold-atomic platforms which are ideally suited to study the thermalization of multimagnon states.
We simulate the nonequilibrium dynamics of these chains using the matrix product state based time-evolution block decimation algorithm and explore the dependence of revivals and thermalization on magnon density and easy-axis onsite anisotropy (which controls the strength of effective magnon interactions).
We observe behaviors akin to those reported for many-body quantum scars which we explain with an analytic approximation that is accurate in the limit of small anisotropy.
\end{abstract}

\maketitle

\section{Introduction}
\label{sec:intro}

\para{}
How does one characterize the low energy spectrum of a system of a large number of electrons?
In many cases, we are fortunate to afford a description of these seemingly complicated many-body systems in terms of quasiparticles.
A comprehensive understanding of their dynamics and the nature of the effective interactions between them provides us with rich information on real material properties at the microscopic level.
Quasiparticles can exist in various forms:
For example, effective electrons in a Landau-Fermi liquid~\cite{sachdev_2011}, spinons in a quantum spin liquid~\cite{balents2010SL}, or magnons in a system with conventional magnetic order~\cite{Henleyobd,Chernyshev_magnon}.

\para{}
Magnetic spin chains provide particularly illuminating examples of quasiparticle physics associated with many deep insights into how strongly correlated electrons collectively act.
In the antiferromagnetic $S=1/2$ chain, for example, neutron scattering sees a continuum of excitations, providing striking confirmation of spin fractionalization and the emergence of spinon quasiparticles~\cite{Faddeev,mourigal2013fractional}, a consequence of correlated many-body effects.
The $S=1$ case is equally spectacular, giving rise to the Haldane spin gap associated with effective $S=1/2$ fractionalized degrees of freedom which are deconfined~\cite{haldane2016ground,haldane1983nonlinear,haldane1983continuum,affleck1989quantum,white1993numerical,Pollmann}.
The low dimensionality of spin chains makes magnetic order highly unstable to quantum fluctuations,
and it is now known that even in higher dimensional systems, geometrical effects such as frustration can achieve similar qualitative outcomes~\cite{balents2010SL,plumb_2018_np}.
In the case of ferromagnets, more conventional magnon quasiparticles are expected and observed, which are well described within the framework of spin wave theory.
Interactions between magnons can lead to distinct signatures in the excited state spectrum~\cite{Chernyshev_magnon}, and corresponding finite frequency observables such as the dynamical magnetic susceptibility.
Adding to the richness of possible emergent behaviors from quasiparticle interactions is the effect of temperature~\cite{Damle_Sachdev,chauhan2019probing},
which must be accounted for to connect to real experiments.
Thus, spin chains provide an ideal setting for exploring the dynamics and effective interactions of magnons and their impact on measurements.

\para{}
$S=1$ systems with predominantly Heisenberg interactions offer an interesting ground for exploring the physics of interacting magnons.
For $S=1$ and higher, terms such as the biquadratic interaction and on-site anisotropy are allowed~\cite{Papanicolaou}, both of which are forbidden for the $S=1/2$ case.
There is a large class of materials and associated realistic models with high spin (see for example~\cite{chauhan2019probing,dally2020three,niesen2011antiferromagnetic,venkatesh2020magnetic,kudasov2006steplike,golinelli1992dispersion, Lauchli_Mila_Penc, plumb_2018_np, zhang_2019, Changlani2015, Sule_2015, Paul2020});
with the ability to perform accurate measurements and theoretical simulations of these systems, there is renewed interest in their physics.
In addition to the plethora of high spin compounds on the materials front, the ability of ultracold atom systems to realize effective high spin models is an exciting opportunity to explore high spin physics in new regimes~\cite{chung2021tunable,jepsen2020spin}.

\para{}
Our work here is inspired by, but not limited to, recent terahertz (THz) optics experiments on NiNb$_2$O$_6$ \cite{chauhan2019probing},
and recent realizations of $S=1$ magnets with tunable anisotropy in ultracold atomic settings~\cite{chung2021tunable}.
In the former experiment, the interaction between magnons was effectively tuned between attractive and repulsive by changing the direction and the strength of the 
external magnetic field (longitudinal vs transverse).
This manifests itself as a significant shift in the location of the excitation in the dynamical susceptibility which moved lower or higher in energy depending on the field direction and the temperature.
More recently, higher-order magnon bound states arising out of magnon interactions have been seen in FeI$_2$ both in neutron~\cite{bai2021hybridized}
and THz optics experiments~\cite{legros2020observation}.
These findings suggest the importance of magnon-magnon interactions that arise from anisotropic terms in the Hamiltonian, which are consequences
of spin-orbit coupling.

\begin{figure}\centering%
	\includegraphics[width=\linewidth]{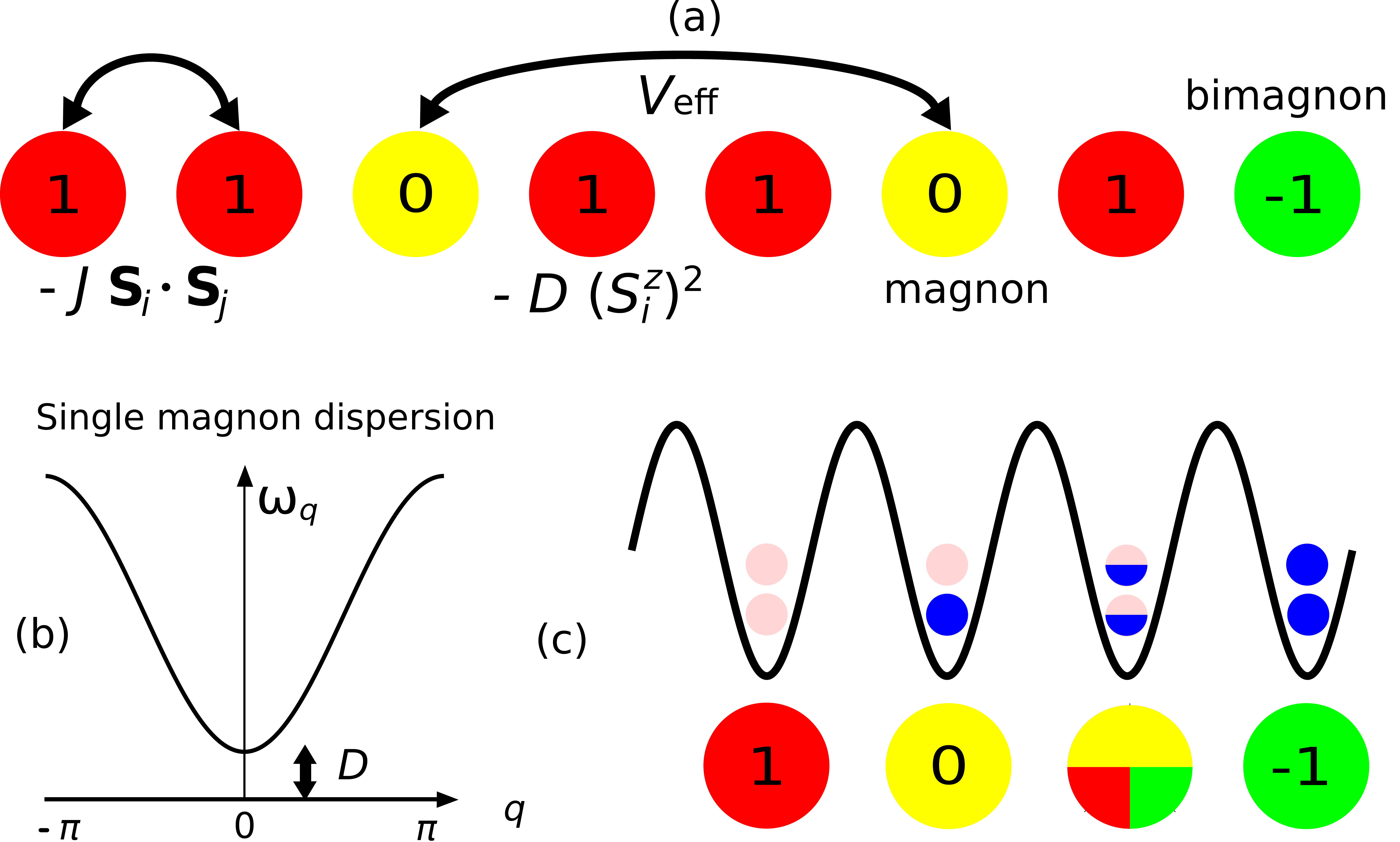}
\caption{\label{fig:single-magnon-dispersion}
(a)
Schematic of a $S=1$ chain showing the terms in the Hamiltonian.
A representative configuration in the Hilbert space is shown.
The ground state corresponds to all sites in the $|1\rangle$ state (or all $|\mathord{-}1\rangle$).
For this ground state, a $|0\rangle$ corresponds to a single magnon and $|\mathord{-}1\rangle$ a bimagnon.
The magnons attract each other via effective interactions mediated by the onsite anisotropy.
(b)
Single magnon dispersion curve for the $S=1$ FM spin chain obtained from linear spin wave theory.
Anisotropy gaps out the lowest energy excitation.
(c)
Mapping between hyperfine states and $S=1$ degrees of freedom as used in recent ultracold atomic setups.
The pink and blue states correspond to hyperfine states of $^{87}\textrm{Rb}$, and the chemical potential is adjusted to two bosons per site.
A microwave pulse can be used to prepare a superposition of hyperfine states, which in turn translates to a superposition of $|1\rangle$,$|0\rangle$, $|\mathord{-}1\rangle$ (depicted with red, yellow and green colors).
More details are in the text.
}
\end{figure}

\para{}
In this paper, we focus on a simple $S=1$ Hamiltonian of direct experimental relevance, both from the point of view of real materials and cold atoms,
\begin{align}
  H &=
	- J \sum_{\langle i,j\rangle } {\textbf{S}_i \cdot \textbf{S}_j}
    - D \sum_{i} (S_i^{z})^2,
\label{eq:spin1ham}
\end{align}
where $J>0$ is the ferromagnetic exchange interaction, $D$ ($>0$ in this paper) is the local onsite uniaxial anisotropy, $S_i^{\mu}$ for $\mu=x,y,z$ are spin-1 operators at site $i$.
This Hamiltonian is schematically depicted in Fig.~\ref{fig:single-magnon-dispersion}(a).
We generally focus on the case of $D \le J$, for NiNb$_2$O$_6$, $J \approx 14.8$\,K (0.308\,THz) and $D \approx 5.2$\,K (0.108\,THz)~\cite{chauhan2019probing}, i.e., $D/J \approx 0.35$.
These parameters correspond to a two-fold ferromagnetic ground state, with the symmetry broken states being either $|1,1,\ldots,1\rangle$ or $|\mathord{-}1,\mathord{-}1, \ldots, \mathord{-}1\rangle$.
Working with one of these ground states as our vacuum, the elementary excitation is a magnon which has well-defined energy and momentum.

\para{}
The questions we pose here are the following: What happens when there are multiple magnons in the system, as is expected at 
finite temperature or with the introduction of a transverse magnetic field? How do magnons interact with one another, 
and what imprint does anisotropy and magnon density leave on the physics of thermalization of multimagnon states?
What is the nature of the composite bound states of magnons, and how do their energetics affect what is observed in the time domain?


\para{}
With these objectives in mind, the paper is organized as follows.
In Sec.~\ref{sec:twomag}, we visit the case of two magnons, and use a combination of the $t$-matrix method, the density matrix renormalization group (DMRG) algorithm~\cite{white1992}, and previously known exact results for two-magnon bound states.
In Sec.~\ref{sec:highermag}, we transfer our lessons to the case of higher magnon bound states, and study the nature of magnon clouds using appropriately defined correlators.
A simple Jastrow function captures all our numerical results surprisingly well, using which we provide both quantitative and qualitative characterization for magnon interactions.
We then focus on the formation of bimagnons in the magnon clouds, and how they grow once they are completely saturated.
In Sec.~\ref{sec:kubo}, we carry forth the acquired insights to address the findings of finite temperature dynamical measurements in NiNb$_2$O$_6$.
In Sec.~\ref{sec:scar} we study the effect of magnon-magnon interactions in the time domain, focusing on the protocol used in $S=1$ cold atom setups~\cite{chung2021tunable}.
We simulate the nonequilibrium dynamics of these chains using the time-evolution block decimation (TEBD) algorithm~\cite{vidal_2003_prl}.
We study revival and thermalization behaviors many of which resemble those seen for quantum many-body scars.
We conclude by summarizing our findings and discussing avenues for possible future experiments.

\section{Two magnon problem}
\label{sec:twomag}

\begin{figure*}
\centering%
\subfigure[]{\includegraphics[height=137pt]{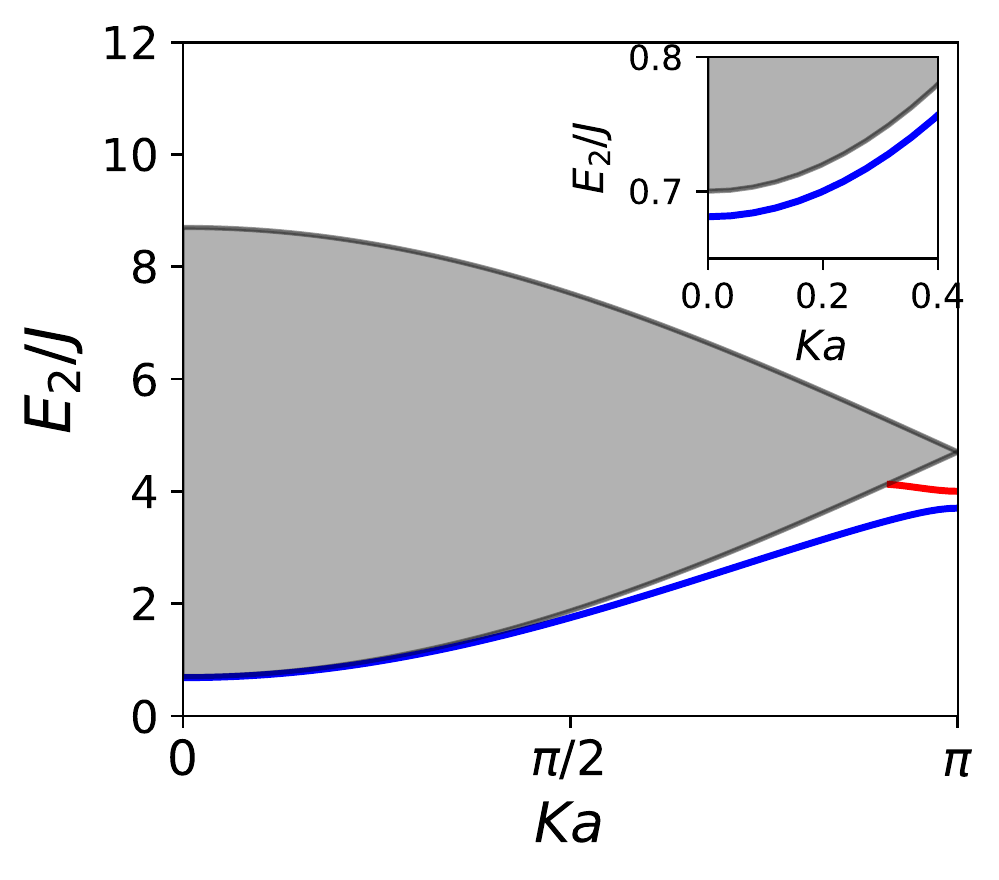}}
\subfigure[]{\includegraphics[height=134pt]{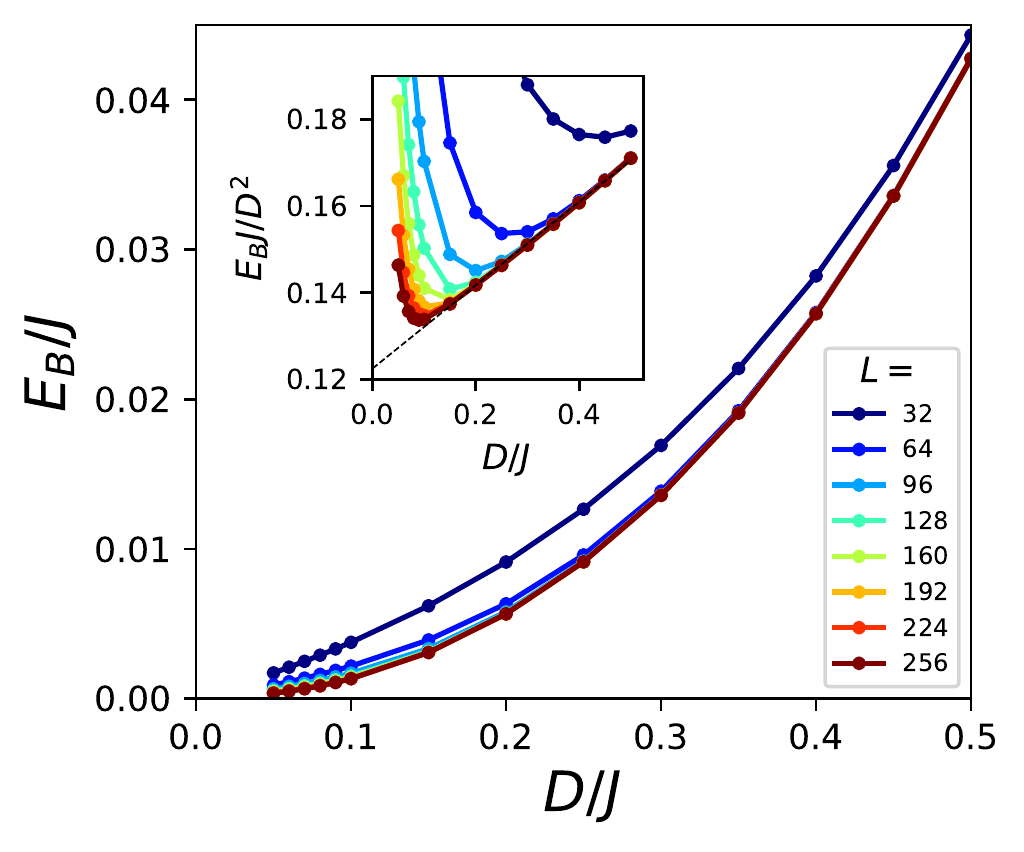}}
\subfigure[]{\includegraphics[height=137pt]{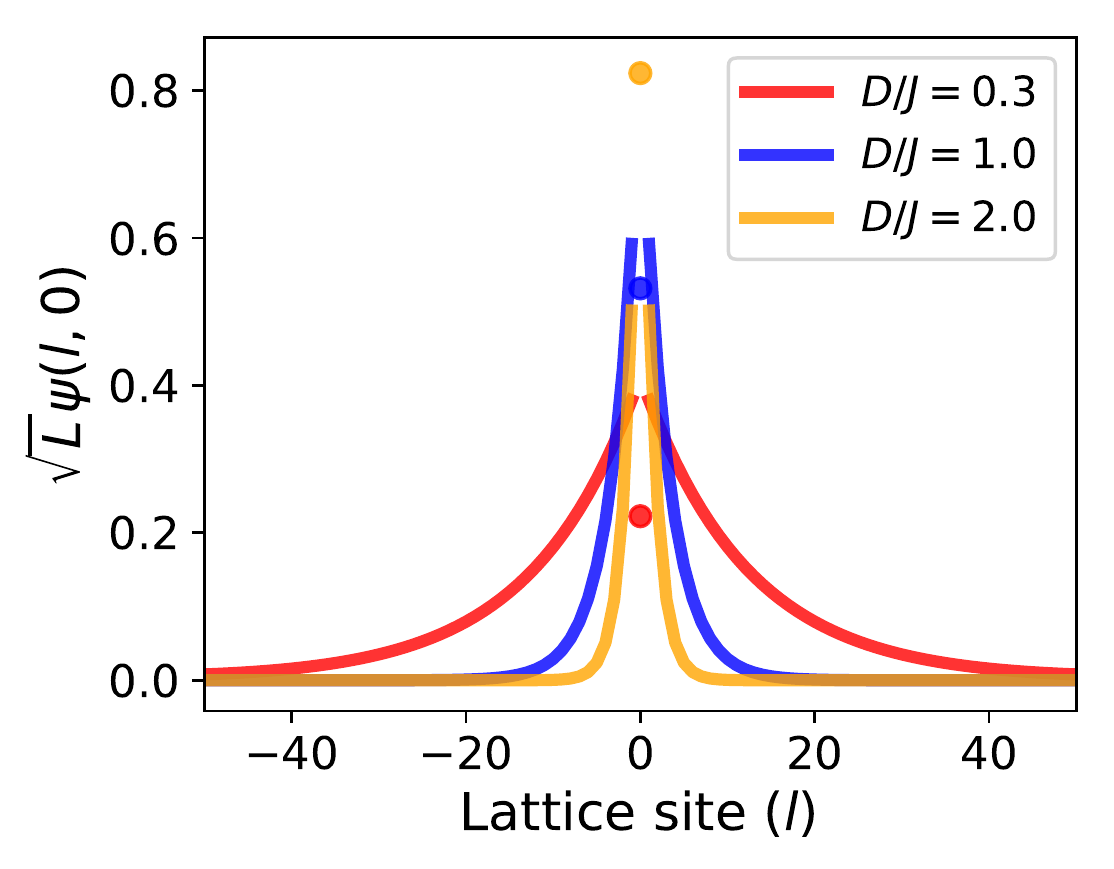}}
\caption{\label{fig:2magdisp}
(a)
Exact two-magnon dispersion for $D/J=0.35$, which is experimentally relevant.
The shaded region indicates the two magnon energy continuum, and the two branches (red and blue) below the continuum represent the two stable bound states.
The lower (blue) branch represents the `Bethe type' bound state while upper (red) branch represents `Ising type' bound state.
At $K=0$, the two-magnon bound state energy lies below the continuum with energy difference $0.019$, which is visible in the inset zoomed over a small region around $K=0$ point.
(b)
Binding energy $E_B/J$ as a function of anisotropy $D/J$ calculated for the lowest-energy two-magnon states.
The inset shows $E_BJ/D^2$ vs $D/J$.
The black dashed line in the inset is a linear fit to the data points excluding the upturn.
The $y$ intercept $\sim 0.122$ agrees with the $t$-matrix result that the leading order term in $E_B$ is quadratic in $D$ with a coefficient of $1/8$.
The size of the bound state increases with decreasing $D$, and the upturns of the curves in the inset shows finite-size effects.
(c)
Probability amplitude of two magnons in the lowest two magnon energy state with one magnon fixed at a reference site `0'.
The circular dots at central site `0' represent the probability of finding two magnons on the same site (i.e., a bimagnon $|\smone \rangle$).
}
\end{figure*}

\para{}
In this section, we review the elementary magnon quasiparticle excitations of the ferromagnetic (FM) spin chain, and use them to build a picture of two magnon bound states.
We characterize properties such as their binding energy and their spatial extent.
Building on previous works, we establish that the uniaxial anisotropy $D>0$ acts as an attractive interaction between magnons, and show that this leads to magnon bound states.

\para{}
For the $S=1$ case a magnon is the lowest energy excitation, which has $S_z=\pm 1$, arising from a spin-flip $\ket{1} \rightarrow \ket{0}$ or $\ket{-1} \rightarrow \ket{0}$.
Since our Hamiltonian in Eq.~\eqref{eq:spin1ham} has total $S_z$ as a good quantum number, magnons can be used as building blocks to describe low energy excitations of various $S_z$ sectors.
Working with the ground state where all sites are in the $\ket{1}$ state, and using the Holstein-Primakoff transformation~\cite{holstein_1940_physrev},
we rewrite Eq.~\eqref{eq:spin1ham} in terms of bosonic magnon creation and annihilation operators $a_{i}^{\dagger}$ and $a_{i}$:
\begin{align}
  S_i^+ &= \sqrt{2S-a_i^\dagger a_i}a_i,
  & S_i^z &= S-a_i^\dagger a_i.
\end{align}
Substituting the above transformations and expanding the Hamiltonian up to quartic order in $a_{i}$ and $a_{i}^{\dagger}$, we get
\begin{align}
  \mathcal{H} &= \mathcal{H}_0 + \mathcal{H}_2 + \mathcal{H}_4 + \ldots.
  \label{eq:hmailtonian-expansion}
\end{align}
where $\mathcal{H}_0 = -LJS^2 - LDS^2$ is the classical ground state energy and $L$ is the total number of sites in the spin chain.

\para{}
The single-particle hopping term $\calH_2$ is given by
\begin{align}
  \mathcal{H}_2 =
    -J S \sum_{\langle i, j \rangle}
      ( a_{i}^{\dagger} a_{j} + \mathrm{H.c.} )
    + \sum_{i} ( (2S-1)D + 2JS ) a_{i}^{\dagger} a_{i},
\end{align}
diagonalizing which gives the bare single magnon (spin wave) dispersion
\begin{align}
  \hbar\omega_{q}
    &=
      2JS \left(
          1- \cos (q a)
      \right)
      +(2S-1) D
\end{align}
where $a$ is the lattice constant of the spin chain.
The schematic for this functional form is shown in Fig.~\ref{fig:single-magnon-dispersion}(b).
The anisotropy term vanishes for $S=1/2$ since the $(S_i^z)^2$ operator cannot distinguish between up and down spins.
The term, however, is allowed for $S=1$, and leads to a gap in the spin-wave dispersion (gives mass to the magnons), stabilizing ferromagnetic order. 
This result for one magnon excited states is exact for the ferromagnetic chain with $S_z$ conservation 
since the higher order terms in Holstein-Primakoff make no contributions to the case of one magnon.

\para{}
We now turn to the case of two magnons.
This problem has been solved exactly for arbitrary spin $S \ge 1$ (in arbitrary dimensions) at zero temperature by Tonegawa~\cite{tonegawa_1970_ptps}.
(Note that the choice of $J$ in Ref.~\cite{tonegawa_1970_ptps} is different from ours by a factor of 2.
The exact results have been adapted to match our convention.)
The key results are as follows.
The energy of a two-magnon bound state in an $S=1$ Heisenberg chain is given as a solution of a cubic equation
\begin{widetext}
\begin{align}
	\left(1+\delta-\epsilon \right)^3 + p_2\left(1+\delta-\epsilon \right)^2 + p_1\left(1+\delta-\epsilon \right) + p_0 = 0,
	\label{eq:exact_cubic}
\end{align}
where the two magnon energy is $E_2 = 4 J   \epsilon$, and
\begin{subequations}
\begin{align}
	p_0 &= -[\{(1-2\alpha)\xi^2 \cos^2(Ka/2)-2\alpha \delta^{'}\}^2 + \delta^{'2}\xi^2 \cos^2(Ka/2)]/{4\alpha},  \\
	p_1 &= -(1-2\alpha-\delta^{'} )\xi^{2} \cos^2(Ka/2) + \delta^{'}(2\alpha + \delta^{'}), \\
	p_2 &= \{(1-4\alpha)\xi^2 \cos^2(Ka/2)-4\alpha(2\delta^{'}+\alpha)\}/{4\alpha}.
\end{align}
\end{subequations}
\end{widetext}
$K = q_1 + q_2$ is the total momentum of the two magnons, $a$ is the lattice constant which we set to 1 for all our calculations,
and the parameters are given by $\xi =1$,
$\delta= \delta^{'} = D/{2J}$, and $\alpha = 1/4$.

\para{}
Even though Eq.~\eqref{eq:exact_cubic} has three solutions, Tonegawa has argued that only two of them are physically meaningful.
Close to the Brillouin zone boundary $Ka = \pi$, both solutions are real-valued, which appear as two branches when plotted as a function of $Ka$.
For one of the solutions, the two magnon-wavefunction has a large amplitude for magnons on nearest neighbor sites,
and smaller amplitude for two magnons on the same site (i.e., a bimagnon corresponding to $|\mathord{-}1\rangle$).
This solution is referred to as the ``Bethe type'' bound state in analogy with the bound state solution first found by Bethe on ferromagnetic $S=1/2$ chains \cite{bethe_1931_zp}.
The other type of bound states are of the ``Ising type,'' 
where the amplitude for two magnons on the same site is larger than the amplitude for them being on neighboring sites.
(This type of bound state is forbidden for $S=1/2$ chains, and hence there is only one bound state branch for the two magnon dispersion in this case).
We note that for $D/J \gtrsim 3.2$~\cite{tonegawa_1970_ptps}, the Ising type two-magnon bound state is lower in energy than the lowest energy single magnon state. 
However, this range of $D/J$ is outside the range explored in this paper, which primarily focuses on smaller $D/J$.

\para{}
Below a certain threshold momentum $K_{\mathrm{th}}$, the cubic equation yields a pair of complex-valued solutions (with nonzero imaginary part), and only one real-valued solution.
The complex-valued solution occurs when the corresponding bound-state branch is no longer stable.
Thus, $K_{\mathrm{th}}$ is the momentum at which the bound state dispersion joins the two particle continuum.
To demonstrate this, we numerically solve Eq.~\eqref{eq:exact_cubic} for $D/J=0.35$ and plot the two-magnon dispersion in Fig.~\ref{fig:2magdisp}(a).
Indeed, we find two branches of two-magnon bound states, with one branch merging with the two-particle continuum.
At $K=0$ we see that the two-magnon bound state is separated from the continuum (see inset).

\para{}
The order of appearance of the two-magnon branches depends on the strength of $D/J$:
The origin of this effect is the touching of the Bethe and Ising branches at $Ka = \pi$. For $D/J \lesssim 0.5$, the Bethe (Ising) type constitutes the lower (upper) branch, and vice versa for $D/J \gtrsim 1.1$.
(This is to be expected:
For small $D/J$, it is energetically unfavorable for two magnons to give up a significant part of their kinetic energy in order to be on the same site,
while at large $D/J$ this bimagnon formation does become favorable.)
For intermediate $D/J$ the behavior is more subtle, for the lowest branch the small $|K|$ wavefunctions are of the Bethe type.
For larger $|K|$, the lower energy branch is of the Ising type, and the higher energy branch is of the Bethe type.
(This subtlety has been discussed in a note in proof by Ref.~\cite{tonegawa_1970_ptps} in response to the study of Ref.~\cite{Torrance}.)
The two branches cross at the zone boundary $Ka=\pi$:
The energy of the Bethe type bound state is given by $3J+2D$ while the energy of the Ising type bound state is fixed at $4J$ and does not depend on $D$.

\para{}
The formation of bound states for $D>0$ can also be captured by treating the quartic order term $\calH_{4}$ in the expansion in Eq.~\eqref{eq:hmailtonian-expansion} that describes the interaction between the magnons, given by
\begin{align}
\mathcal{H}_4
  &=
    - D \sum_i a_i^\dagger a_i^\dagger a_i a_i
    - \frac{J}{2}
      \sum_{\langle i j \rangle}
        a_i^\dagger a_j^\dagger a_j a_i
        \nonumber\\
  &\quad
    + \frac{J}{8}
      \sum_{\langle i j \rangle}
        \left(
          a_i^\dagger a_i^\dagger a_i a_j
          + a_j^\dagger a_j^\dagger a_j a_i
          + \mathrm{H.c.}
        \right),
\end{align}
using the $t$-matrix approach.
The details of our computations are discussed at length in Appendix~\ref{sec:t-matrix}.
Here we highlight the key ingredients of our calculation.

\para{}
The Heisenberg term $J$ contributes to interaction between magnons on neighboring sites,
while the anisotropy term $D$ serves as attractive interaction between the magnons on the same site.
Keeping only the on-site interaction $D$, since the $J$ term vanishes for $q\rightarrow 0$,
the two-magnon susceptibility within the $t$-matrix approximation is
\begin{align}
  \chi_{2}^{t}(q, \omega)
    &= \frac{\chi^{(0)}(q, \omega)}{1 - D \chi^{(0)}(q, \omega)},
\end{align}
where $\chi^{(0)}$ is the Lindhard susceptibility
\begin{align}
  \chi^{(0)}(q, \omega)
    &=
      \int \! \frac{\rmd^d k}{(2\pi)^d}
      \frac{-1}{\omega + i 0^+- (\epsilon_{-k} + \epsilon_{k+q})}.
\end{align}
We define the binding energy of two magnons as
\begin{align}
	E_B \equiv (E_2 - E_0) - 2(E_1 - E_0) = E_2 - 2 E_1 + E_0
\end{align}
where $E_0$, $E_1$, and $E_2$ are respectively the energies of the lowest lying states in the 0-magnon, 1-magnon, and 2-magnon sectors.
At $q=0$, where the two-magnon bound state energy is minimized, we find that $E_{B}$, identified by the location of poles in $\chi^{t}_{2}$, is
\begin{align}
  E_B = \frac{D^2}{8J}.
\end{align}
The result for the ferromagnet is in sharp contrast to bound states in high spin antiferromagnets, where the binding energy was found to be $E_B = \sqrt{JD/2S^2}$~\cite{dally2018amplitude}.
We also note that when the magnon interaction is repulsive, $\chi_2^{t}$ does not allow a pole, and thus no bound state exists.

\para{}
We assess the validity of the above approaches with the help of (almost) exact DMRG calculations for two magnons (in the $S_z=L-2$ sector) for various system sizes.
Fig.~\ref{fig:2magdisp}(b) shows our results from DMRG as functions of the anisotropy $D/J$.
Since the two magnons are only weakly bound for small $D/J$, we observe significant finite size effects in the binding energy in this regime (the upturn).
However, by carefully extrapolating $E_B/D^2$ to the $D \rightarrow 0$ limit, we do find that it approaches a value close to $1/8$, which is consistent with the $t$-matrix results.
Furthermore, Taylor expansion of the exact solution from Ref.~\cite{tonegawa_1970_ptps} is given by
\begin{align}
\frac{E_B}{J}
&= \frac{(D/J)^2}{8}
  + \frac{5 (D/J)^3}{64}
  + \frac{57 (D/J)^4}{2048}
  + \cdots.
\end{align}
This confirms the importance of the on-site interaction provided by $D$ for magnons near the Brillouin zone center, an assumption made in the $t$-matrix approach.

\para{}
Our results are further strengthened by evaluating the scalar components $\psi(l,l')$ (in the notation of Ref.~\cite{tonegawa_1970_ptps}) of the two magnon wavefunction $|\psi_2 \rangle$,
\begin{align}
	|\psi_2\rangle &= \frac{1}{2}\sum_{l'\leq l} \psi(l,l') S_l^{-} S_{l'}^{-} | \psi_{0} \rangle
	\label{eq:tonegawa_psi}
\end{align}
where $l,l'$ are site indices and $|\psi_0 \rangle = \bigotimes_{i=1}^{L}  |1\rangle$ is the ferromagnetic ground state (with zero magnons), and $S^{-}_{l(l')}$ are the spin lowering operators.
We focus on the ground state which we find to be in the $K=0$ sector.
Fixing a reference site $0$, our plot in Fig.~\ref{fig:2magdisp}(c) confirms that for small $D/J$,
the bound state is indeed of the Bethe type ($|\psi(0,0)|<|\psi(1,0)|$),
and for large $D/J$ it is of the Ising type ($|\psi(0,0)|>|\psi(1,0)|$),
with a crossover ($|\psi(0,0)| \approx |\psi(1,0)|$) at $D/J \approx 1.15$.
We will see later that when the magnon number increases, this bimagnon/doublon formation becomes increasingly important even for small $D/J$.

\section{Multimagnon problem}
\label{sec:highermag}

\begin{figure*}
\subfigure[\label{fig:three-magnon-correlator-a}]{\includegraphics[width=\linewidth]{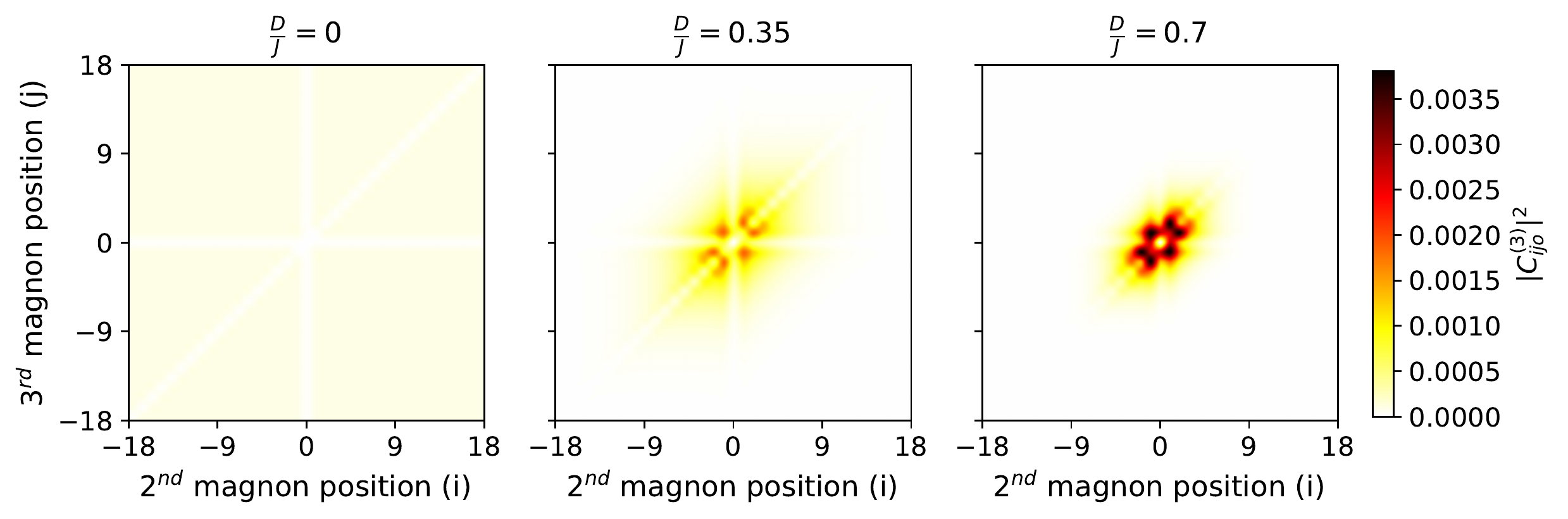}}
\subfigure[\label{fig:three-magnon-correlator-b}]{\includegraphics[width=0.66\linewidth]{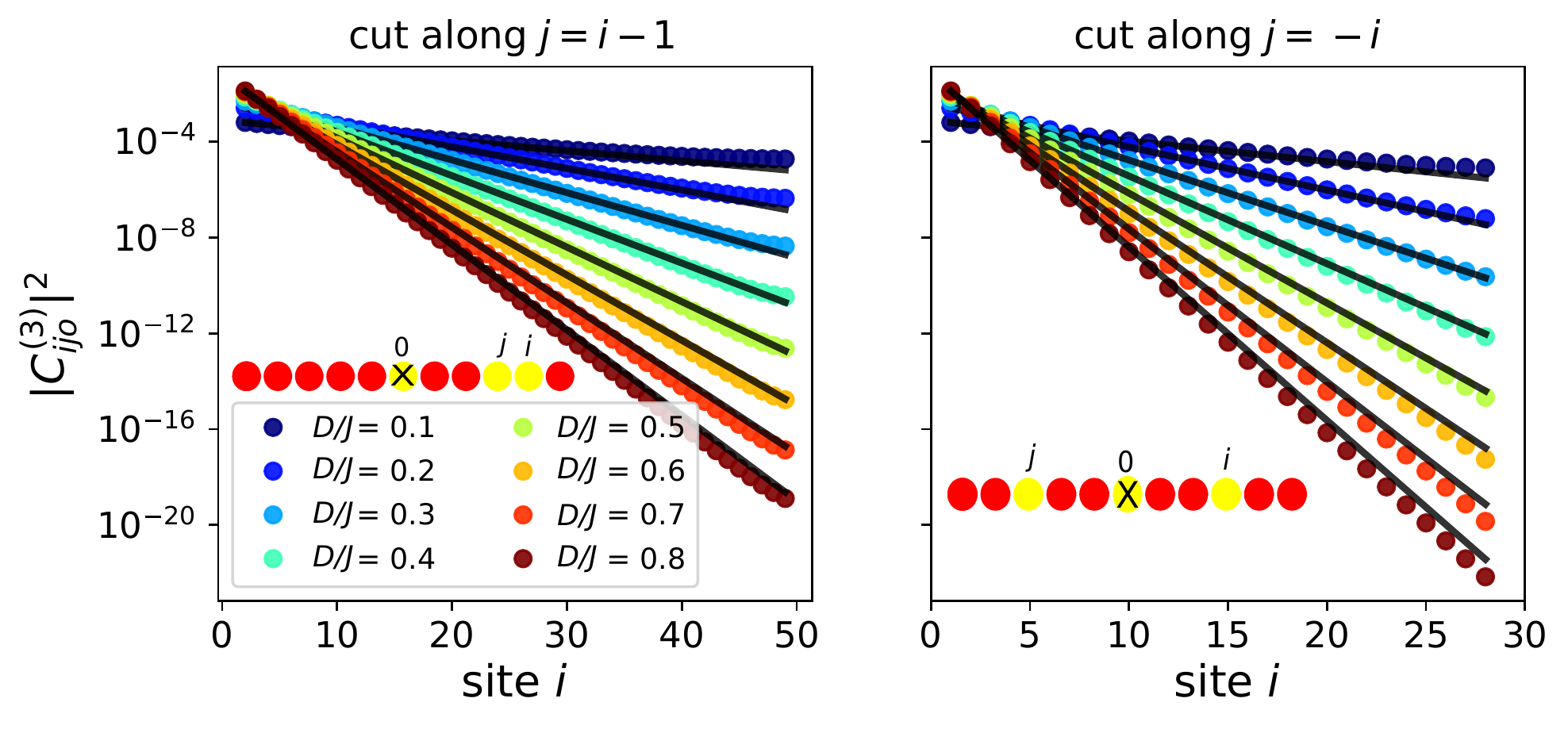}}%
\subfigure[\label{fig:three-magnon-correlator-c}]{\includegraphics[width=0.32\linewidth]{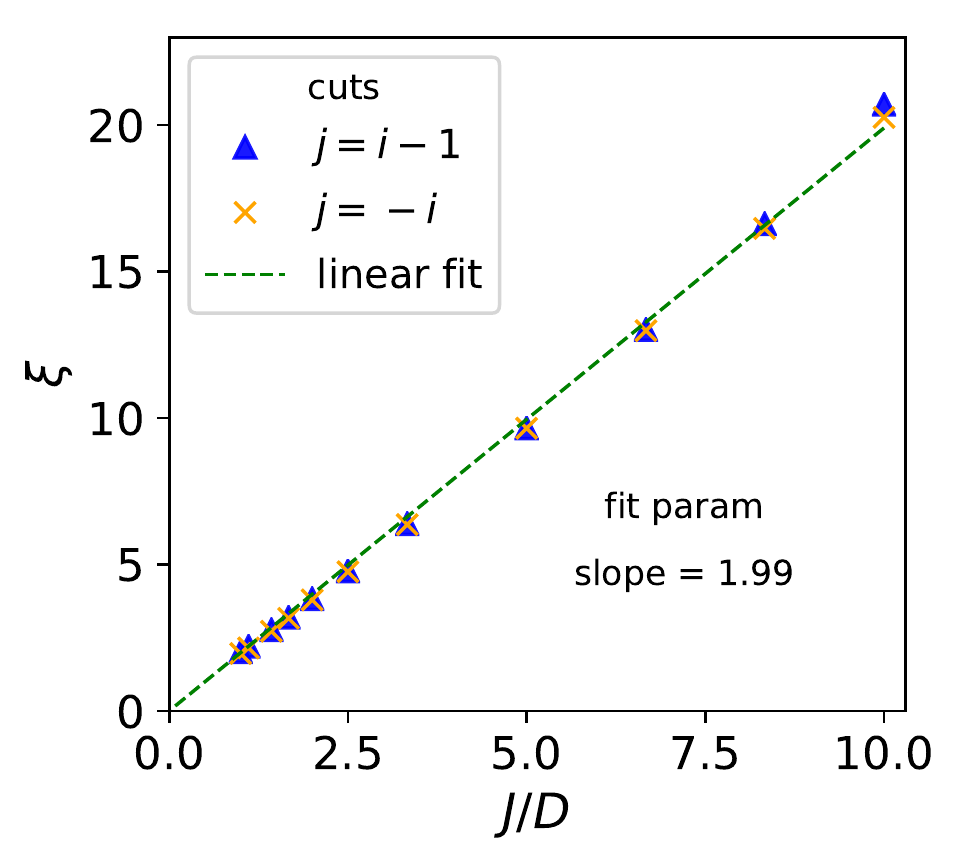}}
\caption{\label{fig:three-magnon-correlator}%
\subref{fig:three-magnon-correlator-a}
The three-magnon correlator $C_{ijo}^{(3)}$ obtained by fixing one magnon at a reference site $o$,
computed for the ground state of three magnons for representative values of anisotropy parameter $D$, in a periodic chain of length $L=40$ using ED.
The colors indicate the magnitude of the correlator.
\subref{fig:three-magnon-correlator-b}
The spatial dependence of the correlator for two one-dimensional cross sections, respectively corresponding to the two magnons at sites $i$ and $j$
with $j=i-1$ (left panel) and $j=-i$ (right panel), while the third magnon is located at $o$.
Both cross sections show exponential decay with increasing spatial dependence between the magnons.
Black solid lines are the fits to the Jastrow wavefunction $\prod_{a<b} e^{|x_a-x_b|/{\xi}}$, with $\xi$ for each cross section being determined independently.
The (almost) exact correlators are calculated for a periodic chain of length $L=100$ with the DMRG algorithm.
\subref{fig:three-magnon-correlator-c}
The two estimates of the correlation length $\xi$ vs inverse anisotropy $J/D$.
The linear fit, shown by the green dashed line, shows excellent agreement with the data.
}
\end{figure*}

\para{}
We now consider the case of more than two magnons.
While the main objective of this section is to build an understanding of their energetics
(which ultimately impacts what is seen in dynamical experiments),
we also explore the qualitative nature of the $n$-magnon ground state wavefunctions.
Specifically, do more than two magnons form bound states? 
If such bound states do form, what are their spatial extent and binding energy? 
Does an additional magnon get assimilated into an existing ``magnon cloud'' or does it break up into magnon molecules? 
Does the formation of single-ion bound states become important during this process?

\para{}
Before proceeding, we briefly review previous work in this direction.
Both Majumdar \textit{et al.}~\cite{majumdar1970application,majumdar1973simple,majumdar1976multimagnon} and \textcite{van1977faddeev} have independently studied three-magnon excitations in Heisenberg ferromagnetic chains using Faddeev's three body formalism~\cite{faddeev2016scattering}.
Although the approach used is completely general, the discussions primarily focused on the $S=1/2$ case.
The study of \textcite{southern1994three} for $S\geq 1$ used the recursion method~\cite{haydock1980recursive} to rigorously argue for the existence of three-magnon bound states based upon the asymptotic behavior of recurrence coefficients and the general features of density of states.

\para{}
We investigate the nature of the lowest energy $n$-magnon state numerically and find that bound states do form for $n \geq 3$, and we study how their properties change with $n$.
For small $n$, an analytically inspired simple Jastrow function captures many of our findings accurately.
This allows us to develop a simple picture for how magnon clouds grow and eventually saturate, which is when bimagnon formation becomes exceedingly important.

\subsection{Ground state wavefunction of \texorpdfstring{$n$}{n} magnons}

\para{}
The wavefunction of $n$ magnons is a superposition of excitations created on top of the ferromagnet,
\begin{align}
	|\psi_n \rangle = \sum_{x_1 \le x_2 \le \ldots \le x_n}
  \!\!
  C(x_1,x_2, \ldots ) S^{-}_{x_1} S^{-}_{x_2} \cdots S^{-}_{x_n} |\psi_0 \rangle
\end{align}
where we have generalized the notation $\psi(l,l')$ from Eq.~\eqref{eq:tonegawa_psi} to the case of $n\ge 2$, with site indices $x_1,x_2, \ldots, x_n$ and absorbed the factor of $2$ from Eq.~\eqref{eq:tonegawa_psi}.
For the case of $S=1$, no three indices can be the same since there is a maximum of two magnons on a given site.
$C(x_1, \ldots)$ represents the amplitude of a particular configuration of magnons, and are complex-valued in general.
Here we focus on the ground state of each multimagnon sector on a periodic chain, which we find to have a net momentum of zero, and $C(x_1, \ldots)$ to be positive.
Instead of evaluating the coefficients, we measure the correlators
\begin{align}
  C_{i_{1} i_{2} ...i_{n-1}o}^{(n)} =\langle\psi_{0}|S_{i_1}^+ S_{i_2}^+... S_{i_n-1}^+ S_o^+|\psi_n\rangle,
\end{align}
where $i_1,i_2,..i_{n-1}$ and $o$ are site indices, with $o$ a fixed reference site.
To do so, we numerically determine 
the lowest energy $n$-magnon state ($|\psi_n \rangle$) using the matrix product state-based DMRG algorithm.

\begin{figure*}
\subfigure[\label{fig:multimagnon-a}]{\includegraphics[width=0.32\linewidth,height=0.21\linewidth]{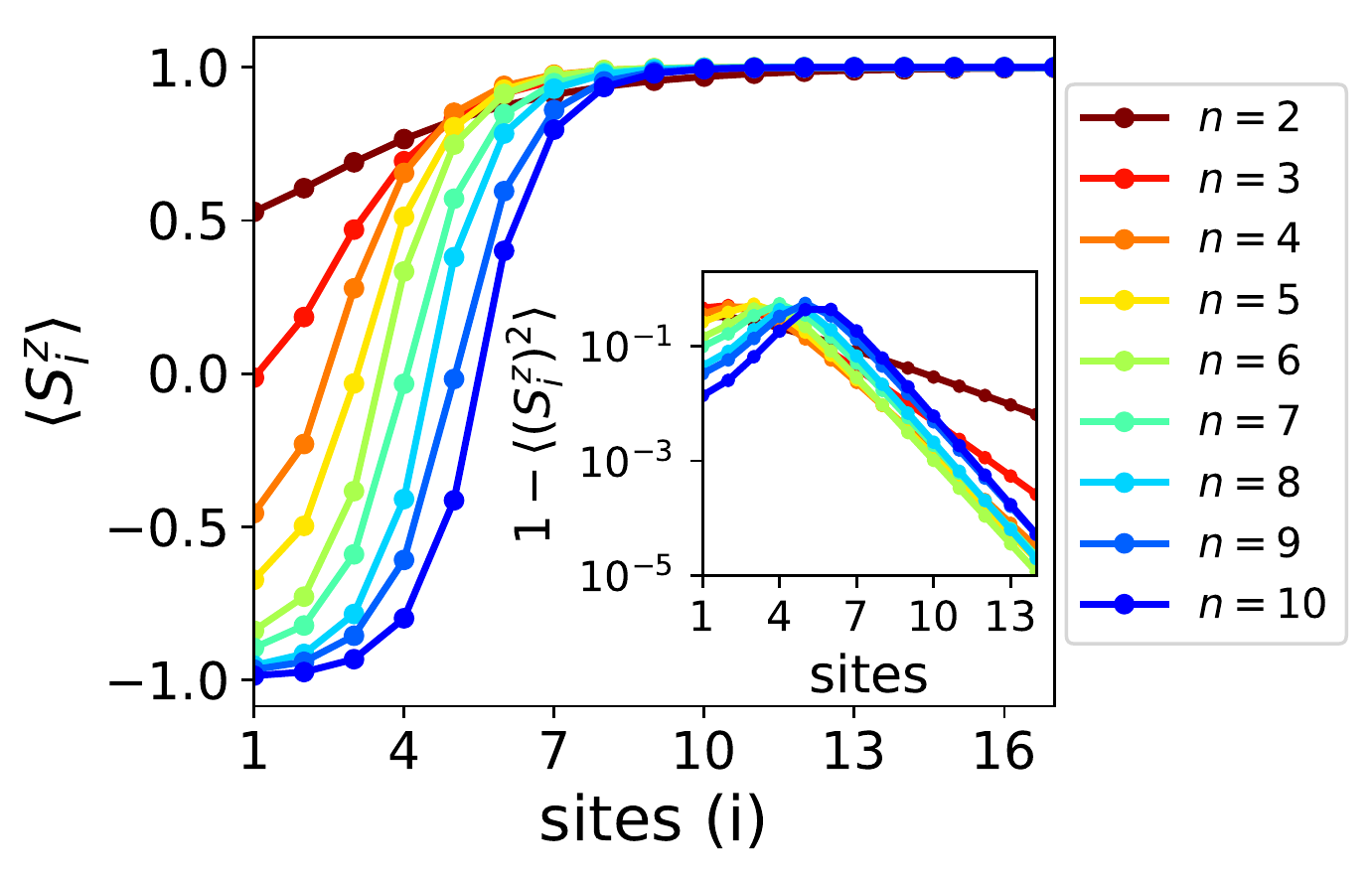}}
\subfigure[\label{fig:multimagnon-b}]{\includegraphics[width=0.27\linewidth,height=0.21\linewidth]{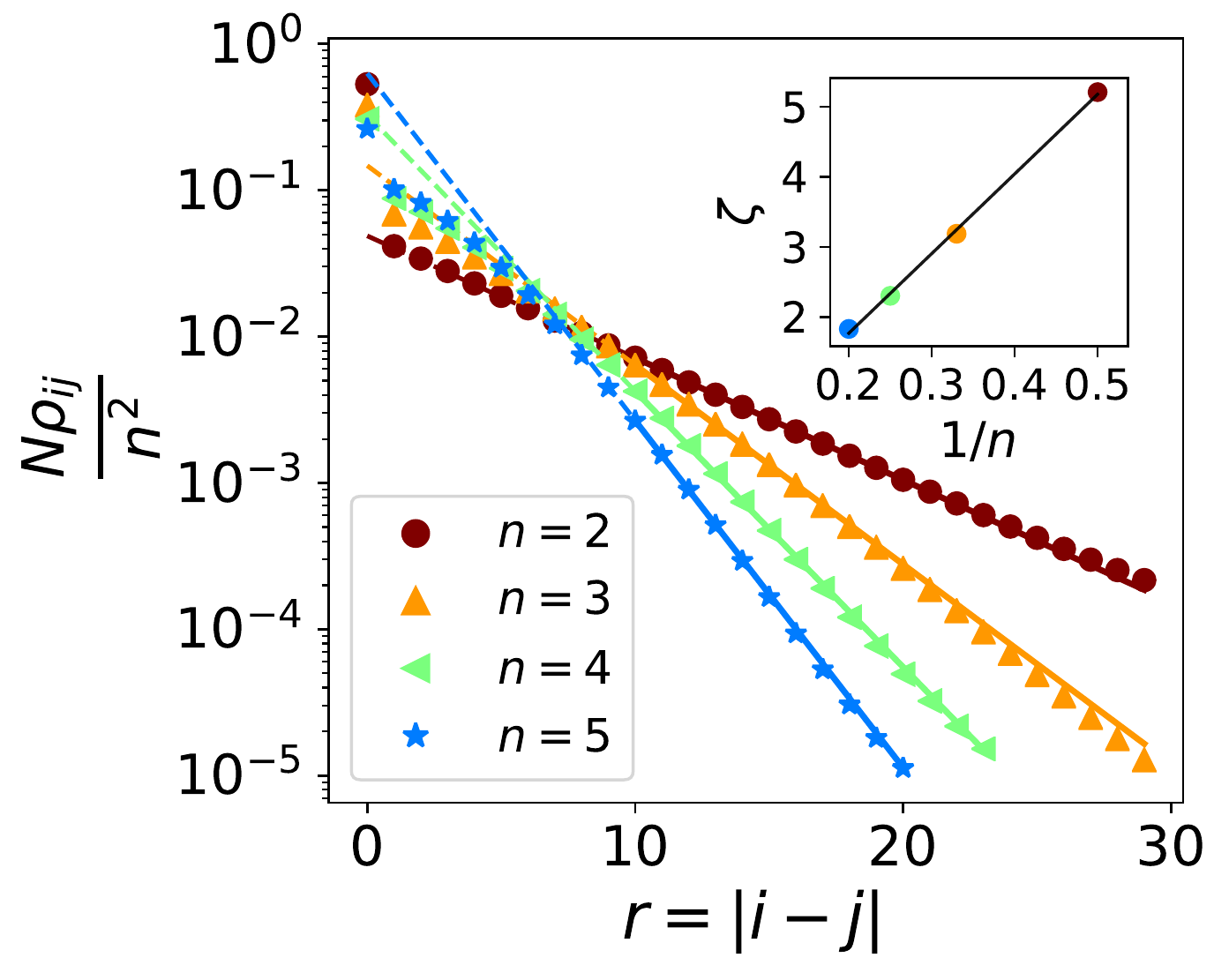}}
\subfigure[\label{fig:multimagnon-c}]{\includegraphics[width=0.27\linewidth,height=0.21\linewidth]{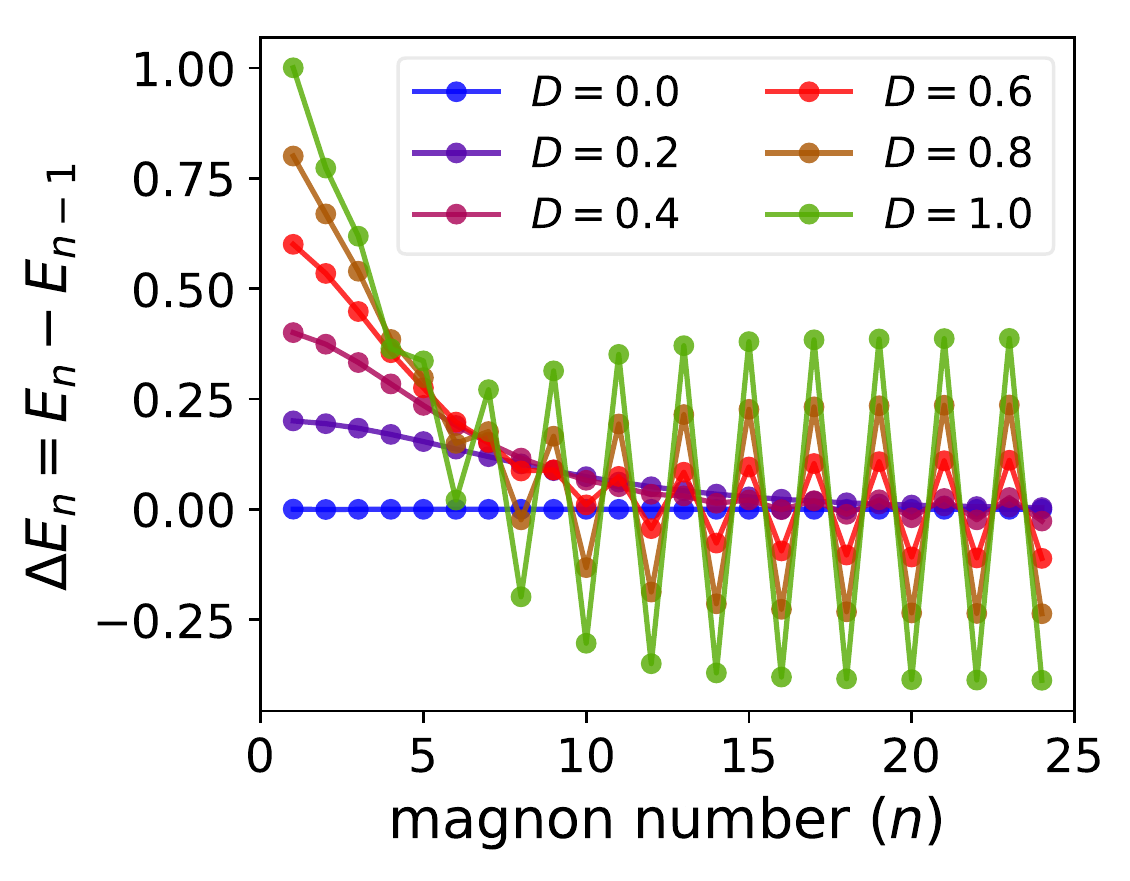}}
\subfigure[\label{fig:multimagnon-d}]{\includegraphics[width=0.115\linewidth,height=0.21\linewidth]{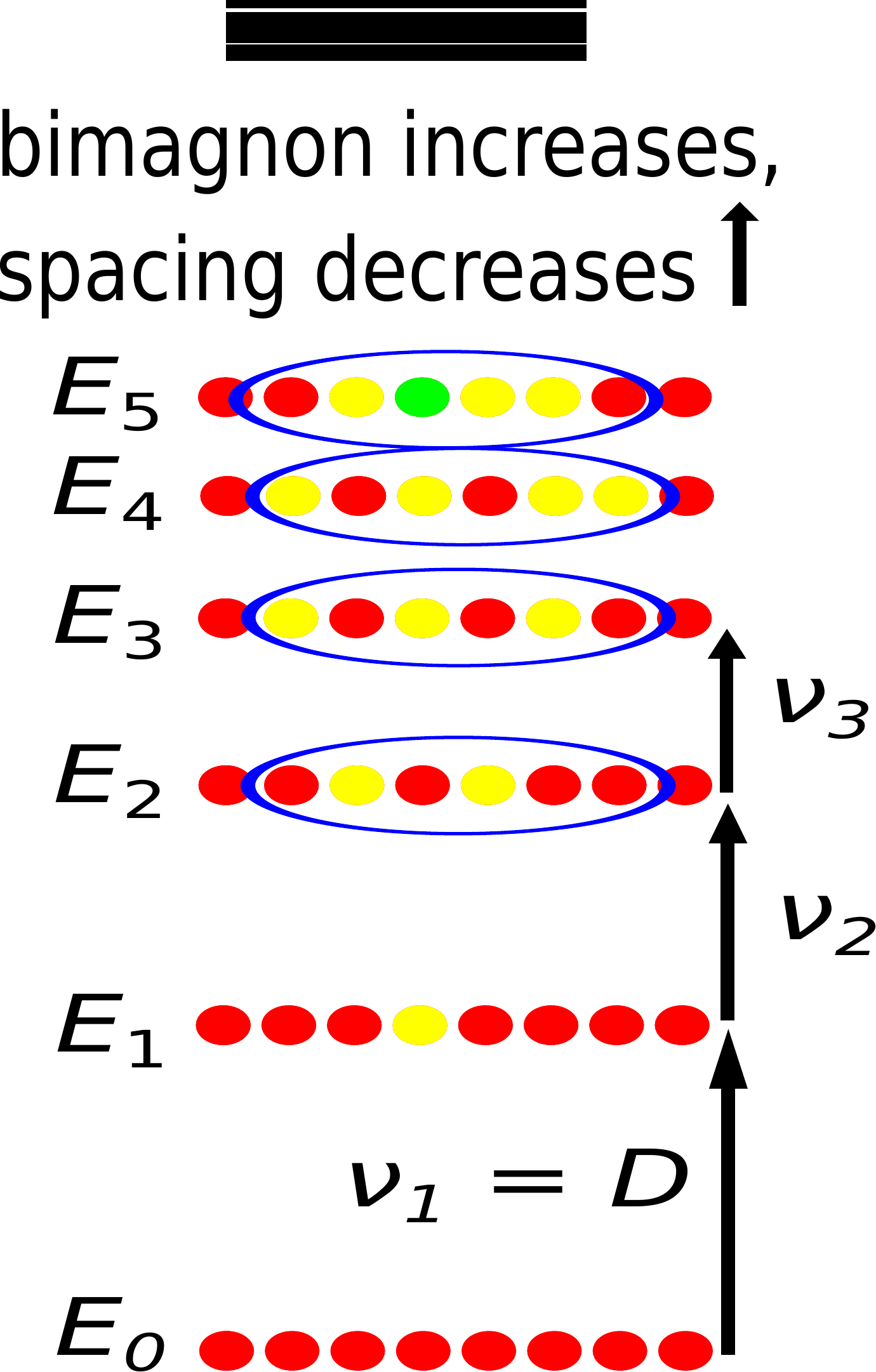}}
\caption{\label{fig:multimagnon}%
\subref{fig:multimagnon-a}
$\langle S_i^z \rangle$ and $1 - \langle (S_{i}^{z})^2 \rangle$ calculated with DMRG for $D/J=0.35$ in the lowest-energy states of various magnon number sectors, on a chain of length $L=100$ with 
open boundary conditions.
The initial states are chosen such that the magnons are on the left end of the chain.
\subref{fig:multimagnon-b}
Normalized pair correlation function $\rho_{ij} \equiv \langle n_{i} n_{j} \rangle$ for $D/J=0.35$ in a periodic chain of length $L=80$ for 2 and 3 magnons, $L=68$ for 4 magnons, and $L=62$ for 5 magnons.
Solid lines are fits to the exponential form $e^{|i-j|/\zeta}$ and  dotted lines are the extrapolations from the long-distance fits.
The inset shows the pair correlation length $\zeta$ vs $1/n$ obtained from the fits.
\subref{fig:multimagnon-c}
Successive energy gaps between consecutive magnon sectors as a function of magnon number for various anisotropy parameters $D$ with $J=1$ fixed in a periodic chain of length $L=180$.
For small $n$, $\Delta E_{n}$ decreases with increasing $n$.
With increasing $n$, oscillatory behavior is observed for $D/J \gtrsim (D/J)_c \approx 0.5$.
\subref{fig:multimagnon-d}
A schematic showing the reduction in transition frequency due to multimagnon bound state formation in different magnon sectors.
}
\label{fig:combined}
\end{figure*}

\para{}
As in the case of two magnons, we find that uniaxial anisotropy $D$ plays an essential role in stabilizing multi-magnon bound states in the spin chain.
Fig.~\ref{fig:three-magnon-correlator-a} shows the three-magnon correlator calculated in the lowest-energy three-magnon state for three representative values of $D$.
For $D=0$, we find that the value of the correlator $C_{ijo}^{(3)}$ is (almost) independent of the $i$ and $j$, suggesting no tendency for the two magnons at these locations to be close to each other or the reference site.
Thus, the ground state shows no hints of bound state formation for $D=0$.
For $D>0$, this picture is dramatically altered, $C_{ijo}^{(3)}$ is now localized around $i=j=o$, and decays with distance.
The magnons clearly cluster together more strongly along the $i\approx j$ line and particularly near $i \approx j \approx o$, the larger $D$ is, indicating bound state formations.

\para{}
These qualitative assertions are made more precise by systematically studying one dimensional cross sections along $j=i+1$ and $j=-i$, as shown in Fig.~\ref{fig:three-magnon-correlator-b}.
The correlator is found to be exponentially decaying with $|i|$ (as may be anticipated);
the precise exponent depends on the cross section under consideration, as we will explain shortly.
For the regime of interest ($D<J$), we find that the amplitude for two magnons on being at the same site (i.e., when two of the indices $i$, $j$, or $o$ are equal) is smaller than the two magnons being at adjacent sites ($i = j \pm 1$), suggesting the magnons form a three-magnon analogue of Bethe type bound state instead of a single-ion type bound state.

\para{}
Based on the above observations, the formation of higher-order bound states (as ground states of the $n$-magnon problem) may be anticipated but is not \textit{a priori} obvious.
This is because there is an inherent competition between the possibility that the system forms a heavy ``droplet'' or ``cloud'' of multiple magnons which stick together, giving up their individual kinetic energy for their collective good,
and the possibility that the magnons split apart into smaller clouds, each of which has its own kinetic energy.
To investigate these scenarios, we carried out calculations for $n=4$, which we report in Appendix~\ref{sec:4magnon}.
We find many qualitative similarities with the cases of $n=2$ and $n=3$, supporting bound-state formation.

\para{}
We now unify our findings to qualitatively and quantitatively understand what happens for $n>4$.
We assert that as long as both the number of magnons and the value of $D/J$ are
sufficiently small, the formation of the single-ion bound state
can be ignored. In this limit, the $n$-magnon wavefunction can be written as a product of pairwise Jastrow factors
\begin{align}
  C(x_1,x_2, \ldots, x_n) \propto \prod_{a<b} e^{-|x_a - x_b|/\xi}.
  \label{eq:jastrow}
\end{align}
In the case of two magnons, $\xi$ is simply the size of the magnon droplet and can be thought of as the correlation length.
For higher number of magnons, it still retains this qualitative interpretation, but the size of the magnon droplet must now be quantified differently to account for the existence of multiple magnons.
We find that $\xi \propto J/D$ to a very good approximation for the case of two, three and four magnons
[see Fig.~\ref{fig:three-magnon-correlator-c} and Fig.~\ref{fig:four-magnon-correlator-b} of Appendix \ref{sec:4magnon}].

\para{}
The power of the Jastrow form becomes most apparent when it is not only used to extract $\xi$,
but also to understand how a collection of magnons organize themselves.
Consider, for example, the case of three magnons with one magnon fixed to the reference site $o$.
According to the Jastrow function, the probability of having two magnons diametrically apart at $j=-i$ is given by the product of three factors
$\exp(-|i|/\xi) \exp(-|i|/\xi) \exp(-2|i|/\xi) = \exp (-4|i|/\xi)$.
In contrast, the probability of having the two magnons at $j = i$ is $\exp (-2|i|/\xi)$.
Thus, the decay length in our fits along the two directions differ by a factor of two.
This is verified by independently fitting the Jastrow function for each cross section at a given $D/J$ and observing perfect consistency between the two estimates of $\xi$, as shown in Fig.~\ref{fig:three-magnon-correlator-c}.

\para{}
For higher $n$, the computation of magnon correlators becomes prohibitive.
For these cases, we estimate the bound state extent by monitoring $\langle S_i^z \rangle$.
The spatial profiles of $\langle S_i^z \rangle$ for open boundary conditions for various $n$ are shown in Fig.~\ref{fig:combined}(a).
With open boundary conditions, the kinetic energy of the magnon cloud is suppressed and it tends to localize spontaneously at either end of the chain:
Here, the cloud is localized at the left end which is seen by the lower value of $\langle S_i^z \rangle$.
On increasing the number of magnons, the value of $\langle S_i^z \rangle$ at the left end approaches $-1$, consistent with the formation of bimagnons.
We complement this information by also plotting the spatial profile of $1-\langle (S_i^z)^2\rangle$ which is presented in the inset of Fig.~\ref{fig:combined}(a).
This metric tells us about the distribution of `0's (i.e., single magnon per site).
The two different spatial profiles show that the formation of single and bimagnons compete with one another.
The formation of bimagnon~($-1$) suppresses the exchange terms (kinetic energy), while $0$'s can hop and lower the energy.
However, there is no single-ion anisotropy cost associated with $-1$ while $0$'s do cost energy. 
This competition is most prominent at the boundaries of the cloud; it does not, however, significantly affect the spatial extent of the cloud itself.
This can be seen in the spatial profile of $\langle S_i^z\rangle$; it saturates to unity roughly at the same length scale ($\sim 10$ lattice constants for the representative value of $D/J=0.35$) for $2$ to $10$ magnons.

\para{}
To shed further light on the properties of the magnon cloud,
we compute the magnon pair correlation
\begin{align}
  \rho_{ij} &= \langle n_i n_j \rangle,
\end{align}
where $n_i$ is the operator that counts the number of magnons at site $i$. 
Substituting $n_i=1-S_i^z$, the pair correlation takes the form (in terms of local spin operators)
$\rho_{ij}=1-\langle S_i^z \rangle - \langle S_j^z \rangle + \langle S_i^z S_j^z \rangle$.
In Fig.~\ref{fig:multimagnon}(b), we plot the pair correlation function for up to five magnons.
We find that the average separation between magnons scales as $1/n$.
This suggests that magnons get closer to each other on average with more magnons in the system.
This is broadly consistent with the cloud being constant in size (for a small number of magnons) with its size being $D$ dependent.
Said differently, as the magnon cloud absorbs additional magnons, it gets heavier and its size does not expand significantly.
However, this cannot continue indefinitely for an arbitrarily large number of magnons, since bimagnon formation eventually saturates the cloud.

\subsection{Energetics of magnon clouds}

\para{}
We now discuss the energetics of multimagnon states.
The energy to introduce an additional magnon into the cloud of $n-1$ magnons is
\begin{align}
  \Delta E_n = E_{n} - E_{n-1},
\end{align}
where $E_n$ and $E_{n-1}$ are the lowest energies of the $n$- and $n-1$-magnon sectors, respectively.
Our calculations for various representative $D/J$, plotted in Fig.~\ref{fig:multimagnon}(c), show that $\Delta E_n$ for small $n$ decreases with $n$, as is expected from the picture that magnons effectively attract one another.
Starting at $\Delta E_1 = D$, $\Delta E_n$ decreases with increasing $n$ up to a $D$-dependent $n$.
Said differently, the energy cost to put an additional magnon into the cloud decreases with increasing $n$.

\begin{figure}
\subfigure[\label{fig:domain-wall-spinprofile}]{\includegraphics[height=114pt]{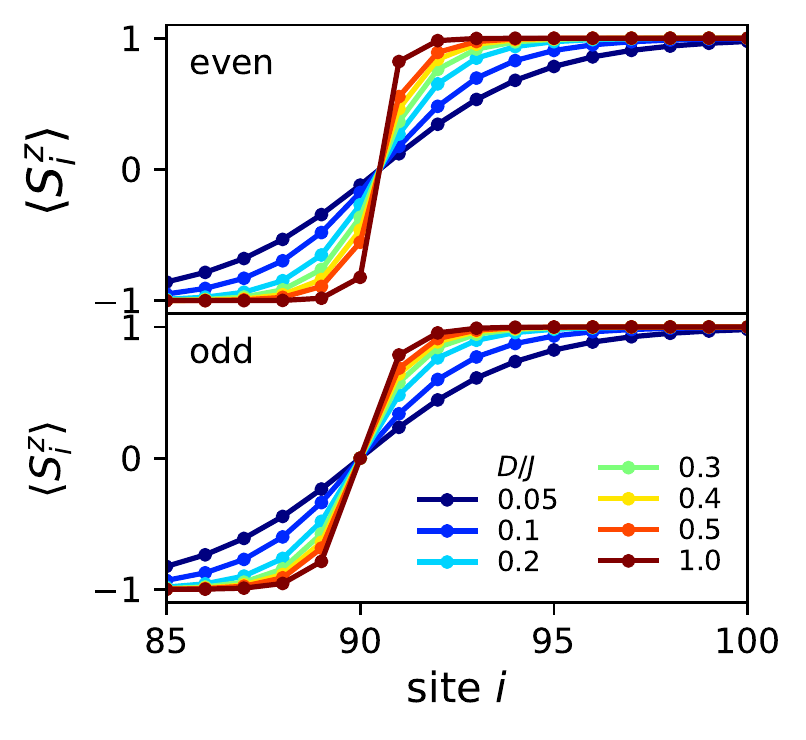}}%
\subfigure[\label{fig:domain-wall-profile}]{\includegraphics[height=114pt]{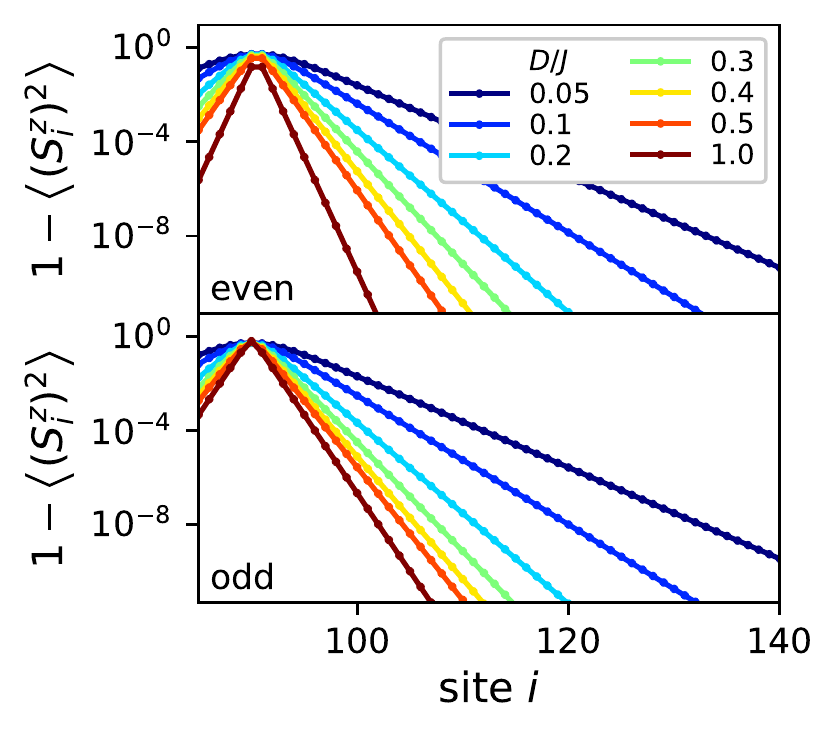}}\\%
\subfigure[\label{fig:domain-wall-thickness}]{\includegraphics[height=114pt]{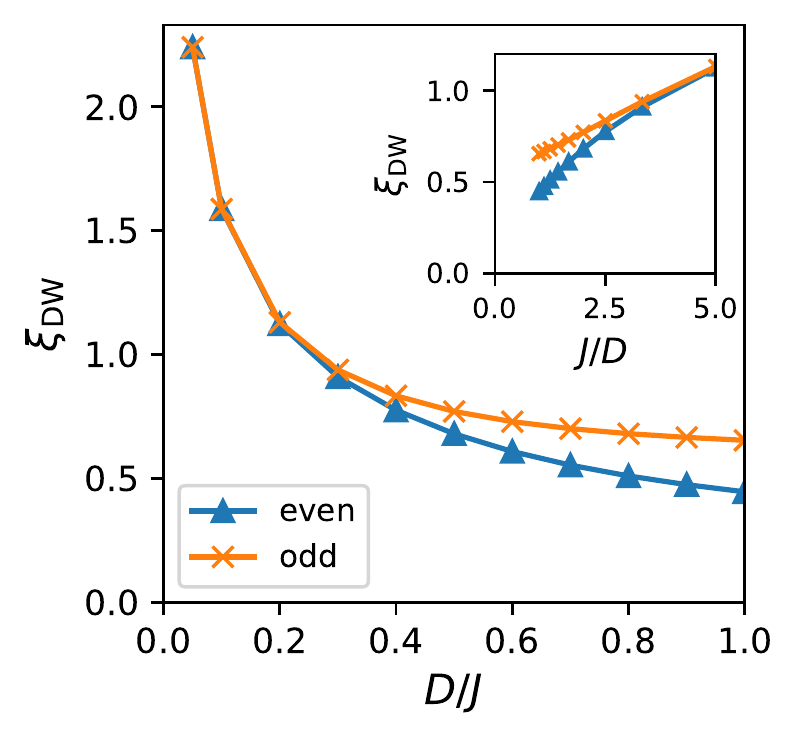}}\;\;\;%
\subfigure[\label{fig:domain-wall-energy}]{\includegraphics[height=114pt]{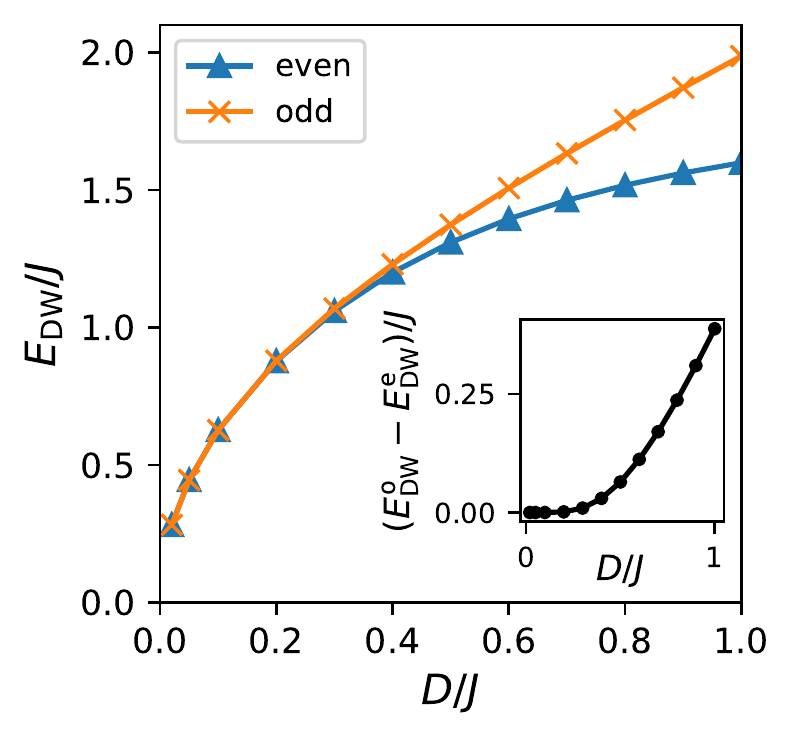}}
\caption{\label{fig:domain-wall}%
\subref{fig:domain-wall-spinprofile}, \subref{fig:domain-wall-profile}
Spatial profiles of the domain walls at various values of $D/J$ for $L=180$ chain.
The labels `even' and `odd' refer respectively to the total $S_z = 0$ and $-1$ sectors.
\subref{fig:domain-wall-thickness}
Thickness $\xi_{\mathrm{DW}}$ of the domain wall vs $D$, from fits to
$1 - \langle (S_i^z)^2 \rangle \sim e^{-|x-x_{\mathrm{DW}}|/\xi_{\mathrm{DW}}}$.
The inset plots the same data as functions of $J/D$,
which shows the scaling $\xi_{\mathrm{DW}} \sim J/D$ for $D \ll J$.
\subref{fig:domain-wall-energy}
Energy of a domain wall $E_{\mathrm{DW}}$ vs $D$.
The inset shows the energy difference between the `odd' domain wall and `even' domain wall.
}
\end{figure}

\para{}
Since the extent of the magnon cloud does not increase appreciably with increasing $n$, it is expected that the Ising type bound states become energetically more favorable than the Bethe type.
When a large number of magnons form a bound state, kinetic energy (XY terms) is highly suppressed, and thus its energetics can be understood from considering the interactions between the magnons (Ising terms).
The $z$ component of Heisenberg exchange and uniaxial anisotropy (i.e., $-J\sum_{\langle i,j \rangle}S_i^zS_j^z$ and $-D\sum_i (S_i^z)^2$), both lower the energy for single-ion Ising like $(|\cdots, \mathord{-}1, \mathord{-}1, \mathord{-}1, \cdots \rangle)$ bound state.
However, both terms contribute nothing to the sites where the Bethe-like $(| \cdots, 0, 0, 0, \cdots \rangle)$ bound state is located.
The even-odd oscillations observed in Fig.~\ref{fig:multimagnon}(c) beyond certain numbers of magnons for large values of $D/J$ are due to the fact that an unpaired magnon ($|0\rangle$) in the odd magnon sectors raises energy because the single-ion anisotropy term favors bimagnon state ($|\smone \rangle$).
Furthermore, the oscillatory behavior is observed for $D \gtrsim 0.5 J$, which coincides with the range of $D$ where the lower energy branch of the two-magnon bound states is of Ising type with significant bimagnon contribution.

\subsection{Dense magnon limit: Domain walls}

\para{}
A complementary picture to the magnons is provided by domain walls (DWs), which captures the behavior of systems with a 
large number of magnons since the two ends of a multimagnon droplet can each be viewed as DWs.
In the Ising limit for $S=1$, there are two types of DWs:
$\ket{\cdots,1,1,\mathord{-}1,\mathord{-}1,\cdots}$ with its center on a bond,
and $\ket{\cdots,1,1,0,\mathord{-}1,\mathord{-}1,\cdots}$ with its center on a site.
Easy-axis anisotropy term prefers the former.
With the inclusion of XY interaction, they acquire thicknesses larger than one lattice constant.
The two types, however, reside in different $S_z$ sectors, and tunneling between one type and the other is allowed only when there are multiple DWs in the system.
The oscillatory behavior in $\Delta E_{n}$, shown in Fig.~\ref{fig:multimagnon}(c), can then be ascribed to the fact that the even magnon sector allows for two bond-centered DWs, which have lower energy, while the odd magnon sector forces one of the DWs to be site-centered.

\para{}
Figures~\ref{fig:domain-wall-spinprofile} and \ref{fig:domain-wall-profile} show spatial profiles of the two types of DWs at various values of $D/J$ on a $L=180$ lattice with open boundary conditions, calculated using DMRG.
A DW here is defined as the lowest energy state in the total $S_z=0$ (for `even') or $-1$ (for `odd') sector, starting from the corresponding configuration in the Ising limit as the initial state.
The DWs have exponential profiles, and their thicknesses decrease with increasing $D/J$ [see Fig.~\ref{fig:domain-wall-thickness}].
At small $D/J$ with thick DWs, the two types of DWs track each other closely;
the two start diverging significantly at $D/J \sim 0.3$, where $\xi_{\mathrm{DW}} \sim 1$ and on-site correlations between magnons become important.
Furthermore, the amplitude of the oscillation in $\Delta E_{n}$ for large $n$ matches the energy difference between the two types of DWs.
[Compare Fig.~\ref{fig:domain-wall-energy} and its inset with Fig.~\ref{fig:multimagnon}(c).]

\begin{figure}
\centering%
\subfigure[\label{fig:kuboAnalysis-a}]{\includegraphics[width=\linewidth]{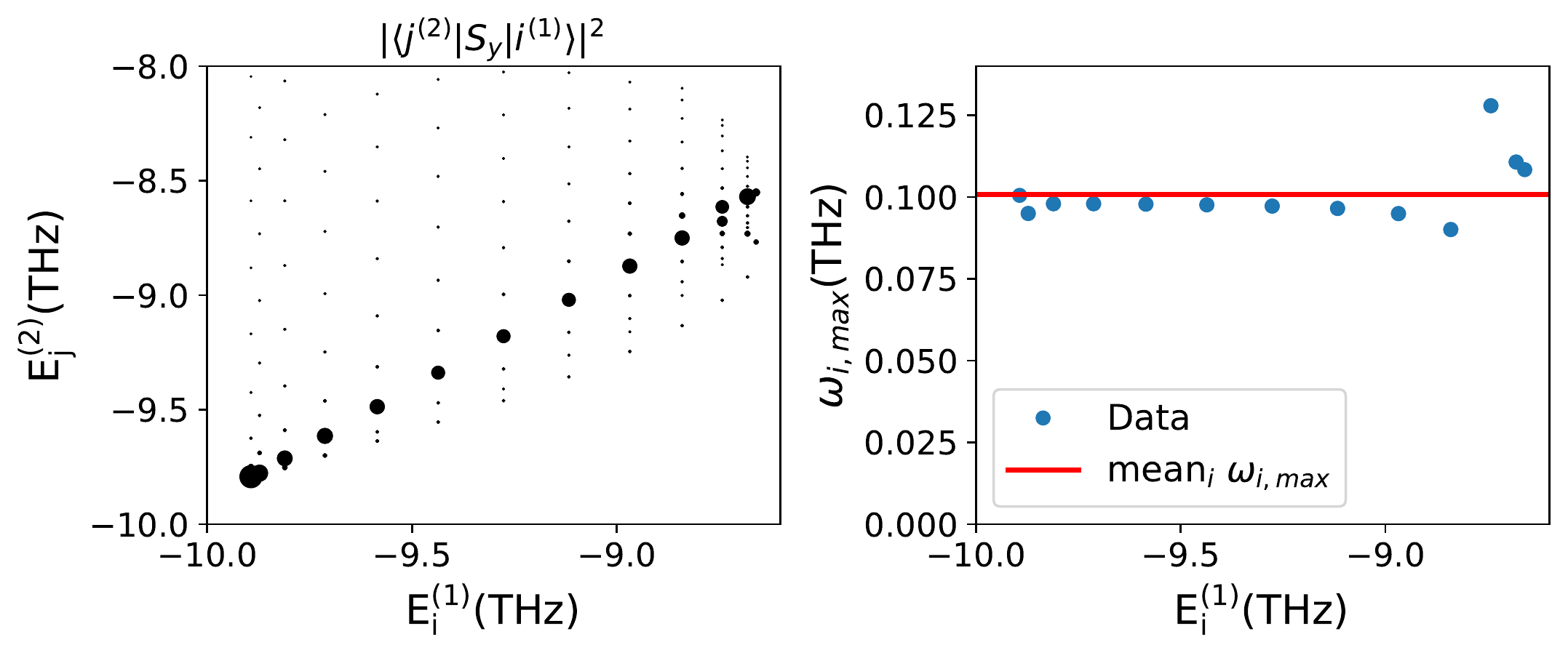}}
\subfigure[\label{fig:kuboAnalysis-b}]{\includegraphics[width=\linewidth]{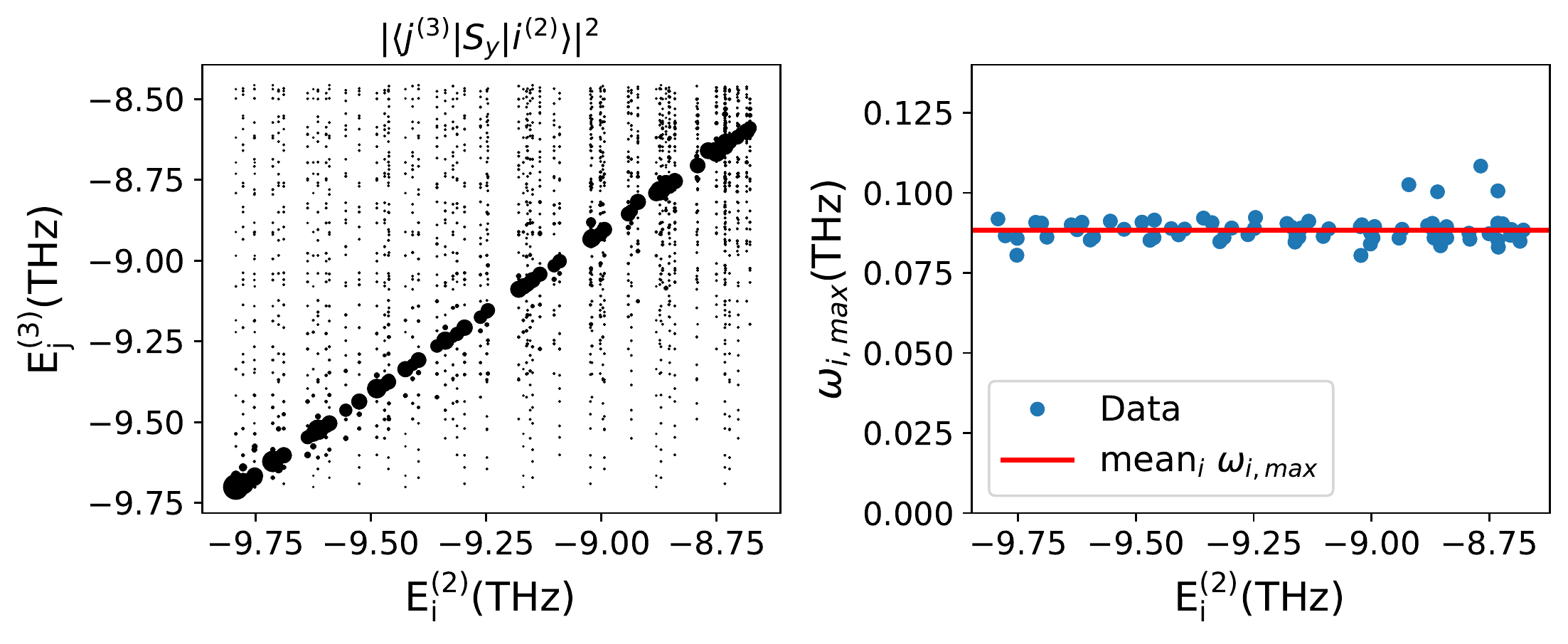}}
\caption{\label{fig:kuboAnalysis}%
Matrix elements $|\langle j^{(n+1)} | S_y | i^{(n)}\rangle |^2$ for transitions between \subref{fig:kuboAnalysis-a} (left panel) for $n=1$ and \subref{fig:kuboAnalysis-b} (left panel) $n=2$ magnon levels calculated in a periodic chain of length $L=24$ using ED and for parameters relevant to NiNb$_2$O$_6$, i.e., $J=0.308$\,THz and $D=0.108$\,THz.
The size of each black dot represents the square of the absolute value of the corresponding matrix element.
Right panels of both \subref{fig:kuboAnalysis-a} and \subref{fig:kuboAnalysis-b} represent the frequency $\omega_{i,\text{max}}$ at which the transitions in the left panels are most active for a given $|i^{(n)}\rangle$,
i.e., corresponding to the $|j^{(n+1)}\rangle$ with the largest absolute value of the matrix element $|\langle j^{(n+1)}|S_y|i^{(n)}\rangle |^2$.
}
\end{figure}

\section{Recap of finite temperature dynamical experiments and connection to our results}
\label{sec:kubo}

\para{}
Till this point, our focus has been on multimagnon states which are ground states of their respective magnon number sectors.
However, an explanation of the finite temperature dynamical susceptibility requires us to develop the connection to excited states (and corresponding matrix elements) which enter the response functions.
In this section, we briefly recap crucial aspects of the THz optics experiment~\cite{chauhan2019probing} on the $S=1$ chain compound NiNb$_2$O$_6$ (which we refer to as the ``JHU experiment") and summarize the key findings.
We build on results presented in earlier sections with the objective of explaining the findings of the JHU experiment.

\para{}
In the JHU experiment, the direction of the chain of the magnetic atoms was referred to as the $z$ axis,
the direction of the incidence of light as the $x$ axis
and the light is linearly polarized with its oscillating magnetic field along the $y$ axis.
(As a first approximation, we will ignore any possible canting of the easy-axis of the spins with respect to the $z$ axis, i.e., the easy-axis anisotropy is perfectly along the 
one-dimensional chain of spins.)
The absorption cross section of linearly polarized light is inferred from the transmission coefficient,
from which the dynamical susceptibility is determined.
The locations of peaks in the dynamical susceptibility reveal information about the energy levels of the system,
allowing indirect inference of which transitions are most active at a given temperature.
In a longitudinal applied field, the most prominent peak in the low frequency susceptibility moves to lower frequency with increasing temperature.
This observation was not reported in CoNb$_2$O$_6$ with effective $S=1/2$ magnetic ions~\cite{morris2014hierarchy}
(albeit with different exchange interactions~\cite{Morris2021,Fava_2020}),
strongly hinting that the $S=1$ nature of the magnetic ions is at the heart of the effect.
Additionally, the direction of the temperature-dependent shift was reported to depend on the direction of the applied static magnetic field.

\para{}
For the case of longitudinal field with strength $B$, magnon number is a good quantum number and all our analyses in the previous sections apply straightforwardly.
The additional Zeeman term $-g \mu_B B S_z$ does not alter the wavefunctions in a given magnon sector, all it contributes is an overall energy shift.
For example, the energy of the one-magnon state (with respect to the ferromagnetic ground state) is given by $\delta_{1}=D+ g \mu_B B$.
More generally, the energy for an additional magnon is $\delta_{n}=(E_{n}-E_{n-1})_{B=0} + g \mu_B B$ where the subscript refers to the corresponding values for the $B=0$ case.
The \textit{shift} in the peak absorption frequency seen in the JHU experiment, on going from low to high temperature, depends only on the \textit{change} $\delta_{n=high} - \delta_{1}$.
This shift is independent of $B$, which is why it is sufficient to analyze only the $B=0$ case to explain its value seen in the experiment.

\para{}
For the case of transverse applied magnetic fields, applied along the $x$ axis, (which we have not considered in this paper),
magnon number is not well defined at small field strengths. For large field strengths $B \gg J,D$,
however, magnon number is approximately conserved on choosing the quantization axis to be along the direction of the applied field.
In this description, magnons mutually repel each other~\cite{chauhan2019probing}.
Due to this repulsion, the peak frequency in the dynamical susceptibility increases with increasing temperature.

\para{}
The JHU experimental findings call for a closer look at the mechanism by which this temperature-dependent energy shift occurs for the $S=1$ chain.
Within linear response theory (Kubo formalism), the dynamical susceptibility at finite temperature is given by
\begin{widetext}
\begin{align}
	\chi_{yy}(\omega,T)=\frac{\pi(1-e^{-\beta \omega})}{Z} \sum_{p,q}e^{-\beta E_p}\left|\langle q|S_y|p\rangle\right|^2 \delta(E_p-E_q+\omega)
	\label{eq:kubo_xyy}
\end{align}
\end{widetext}
where $\omega$ is the frequency being probed,
$T$ and $\beta = 1/k_B T$ are the temperature and inverse temperature, respectively,
$Z=\sum_p e^{-\beta E_p}$ is the partition function,
$|p\rangle$ are the eigenstates of the Hamiltonian with the energy eigenvalues $E_p$, and $S_y=\sum_i S^y_i$.
Therefore, the transition matrix element $\langle q|S_y|p\rangle$ is nonzero only for $|p\rangle$ and $|q\rangle$ that differ in total $S_z$ quantum number by one unit of angular momentum.
Thus, the transition frequency $\omega=E_q-E_p$ is the energy difference that involves states in two consecutive magnon sectors.
(We will attach an additional label to the state label to indicate the magnon number sector it belongs to.)

\para{}
In the previous section, we showed that this energy difference decreases from $D$ to zero with increasing magnon number, followed by even-odd oscillations.
(For $D/J=0.35$ we find these oscillations to be fairly weak.
However, they are significantly strengthened at larger $D/J$, for $D/J \gtrsim 0.5$ which is when the Ising type/bimagnon bound states become important.
This effect is potentially observable in systems where a large anisotropy can be realized.)
Since higher magnon sectors are entropically favored at high temperature, our calculations suggest that the peak frequency should be reduced by an amount of $D$ when the temperature is increased from low temperature $T\ll J,D$ to the high temperature limit $T \gg J,D$.
This observation is consistent with the findings of the JHU experiment for the case of longitudinal fields~(their Fig.~3 for $B=65$\,kG in Ref.~\cite{chauhan2019probing})--the peak in the dynamical susceptibility moves from $0.28$\,THz to $0.20$\,THz on increasing the temperature from $5$\,K to $50$\,K.
Given the simplistic modeling of the spin chain, this observed shift of $0.08$\,THz is in reasonable agreement with the theoretical estimate of $D=0.108$\,THz.

\para{}
The above argument relies on a simplifying assumption that only the transitions 
between the lowest energy states of $n-$ and $n+1-$magnon sectors are important. 
However, according to the Kubo formula in Eq.~\eqref{eq:kubo_xyy},
\textit{all possible} contributions arising from transitions must be accounted for.
$\chi_{yy}(\omega, T)$ measures the appropriately weighted sum of all possible transitions consistent with the selection rules
(no change in linear momentum, and change of spin angular momentum by one quantum), but does not 
provide sufficient information for inferring individual contributions from each eigenstate. 
Numerical exact diagonalization calculations provide this additional knowledge (for small number of magnons), 
which we use in Fig.~\ref{fig:kuboAnalysis} to plot the transition matrix elements between $1 \rightarrow 2$ magnon and $2 \rightarrow 3$ magnon energy levels for 
material-specific parameters. The many-body eigenstates in the $n$- ($n+1$-) magnon sector, organized by increasing energy, are on the horizontal (vertical) axis.
The size of each black dot is proportionate to the matrix element $\left|\langle j^{(n+1)}|S_y|i^{(n)}\rangle\right|^2$ representing the transition between a pair of energy levels $|i^{(n)} \rangle$ (with energy $E^{(n)}_i$ in the $n$-magnon sector) and $|j^{(n+1)} \rangle$ (with energy $E^{(n+1)}_j$ in the $n+1$-magnon sector).

\para{}
The results suggest that not all (symmetry allowed) transitions are equally important:
For each $E^{(n)}_i$, most of the weight is concentrated on a single $E^{(n+1)}_j$, and other contributions are small.
For each $|i^{(n)}\rangle$ we identify the frequency $\omega=E^{(n+1)}_{j}-E^{(n)}_{i}$ for which the matrix element is largest and refer to it as $\omega_{i,\text{max}}$.
For the cases considered ($n=1$ and $n=2$) we find that $\omega_{i,\text{max}}$ is largely independent of $|i^{(n)}\rangle$ in a given $n-$magnon sector.
Importantly, this average/typical value of $\omega_{i,\text{max}}$ (indicated by the red line in each panel) decreases with increasing magnon number.
Although this shift across $1\rightarrow2$ and $2\rightarrow3$ magnon sectors is small, the general trend is consistent with our earlier findings.
Said differently, the effective energy cost to add an additional magnon (with zero additional momentum, as dictated by the matrix element selection rules) \textit{decreases} with \textit{increasing} magnon number.
This holds not only for the lowest energy state of each $n$-magnon sector but also for the \textit{excited states}.

\section{Nonequilibrium dynamics of magnons in ultracold atomic settings}
\label{sec:scar}

\begin{figure*}[]
\subfigure[]{\includegraphics[width=0.49\linewidth]{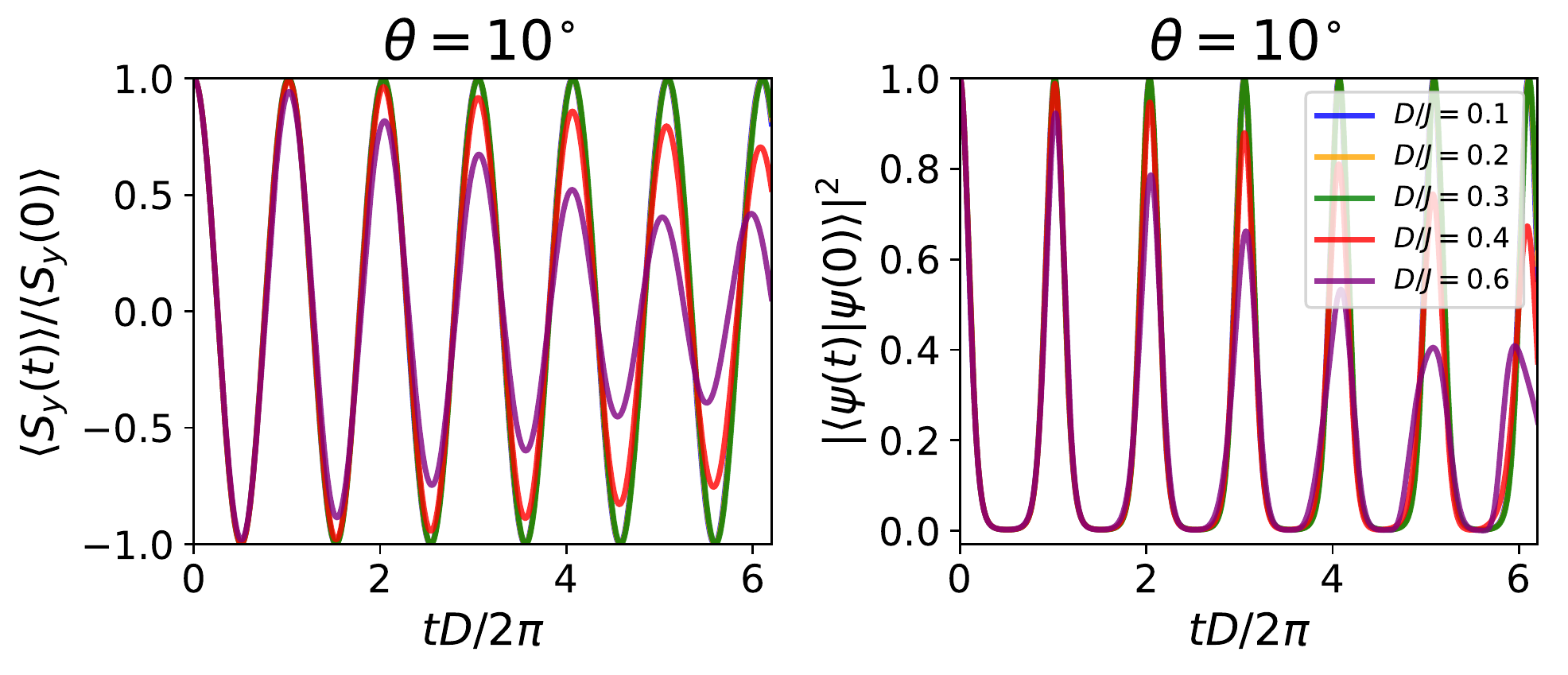}}
\subfigure[]{\includegraphics[width=0.49\linewidth]{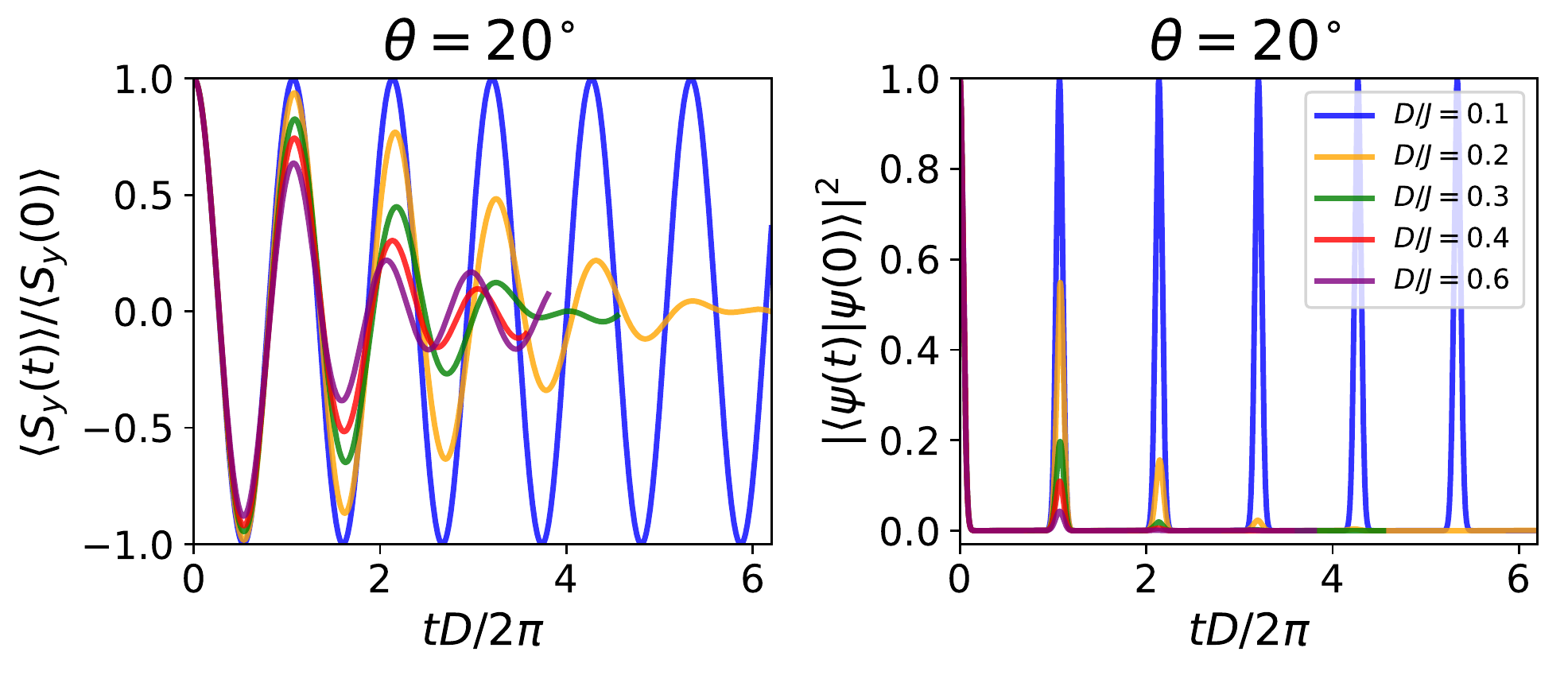}}
\subfigure[]{\includegraphics[width=0.49\linewidth]{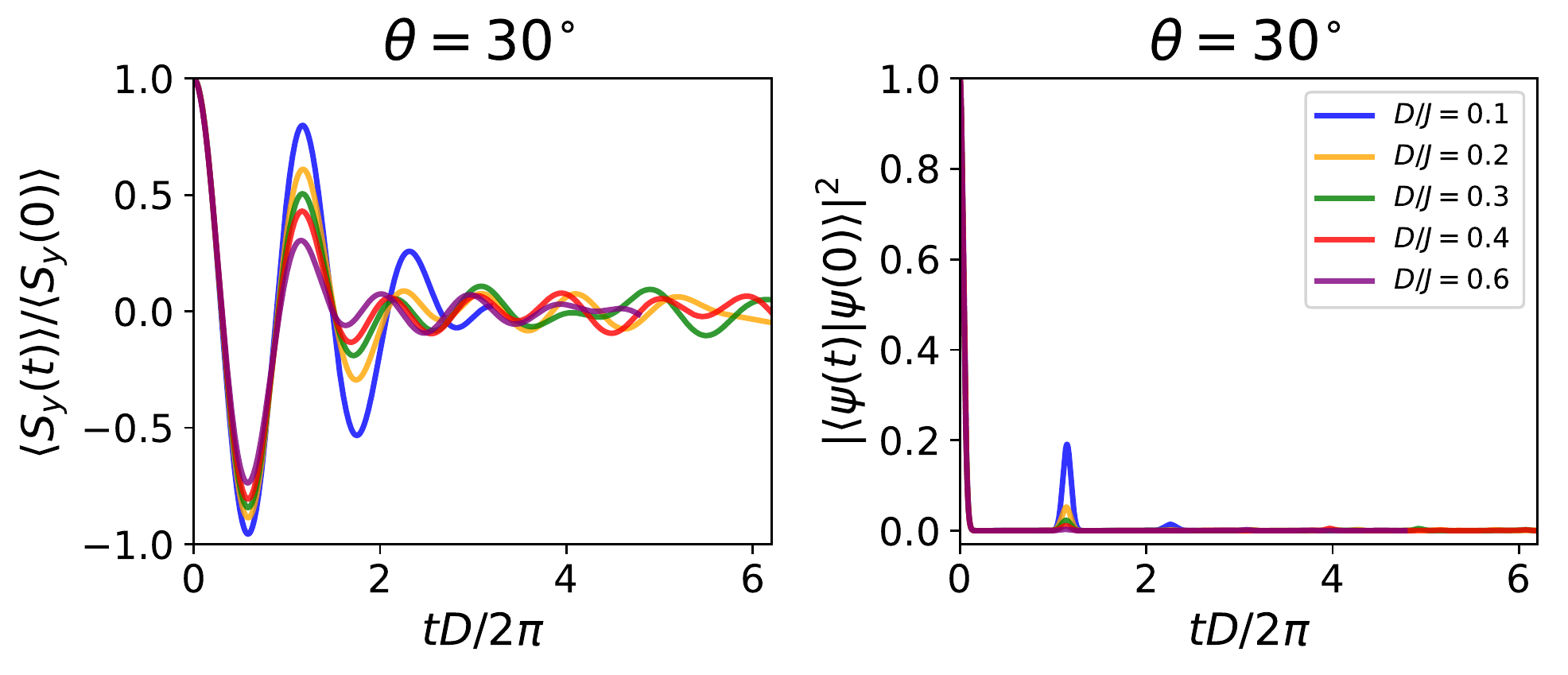}}
\subfigure[]{\includegraphics[width=0.49\linewidth]{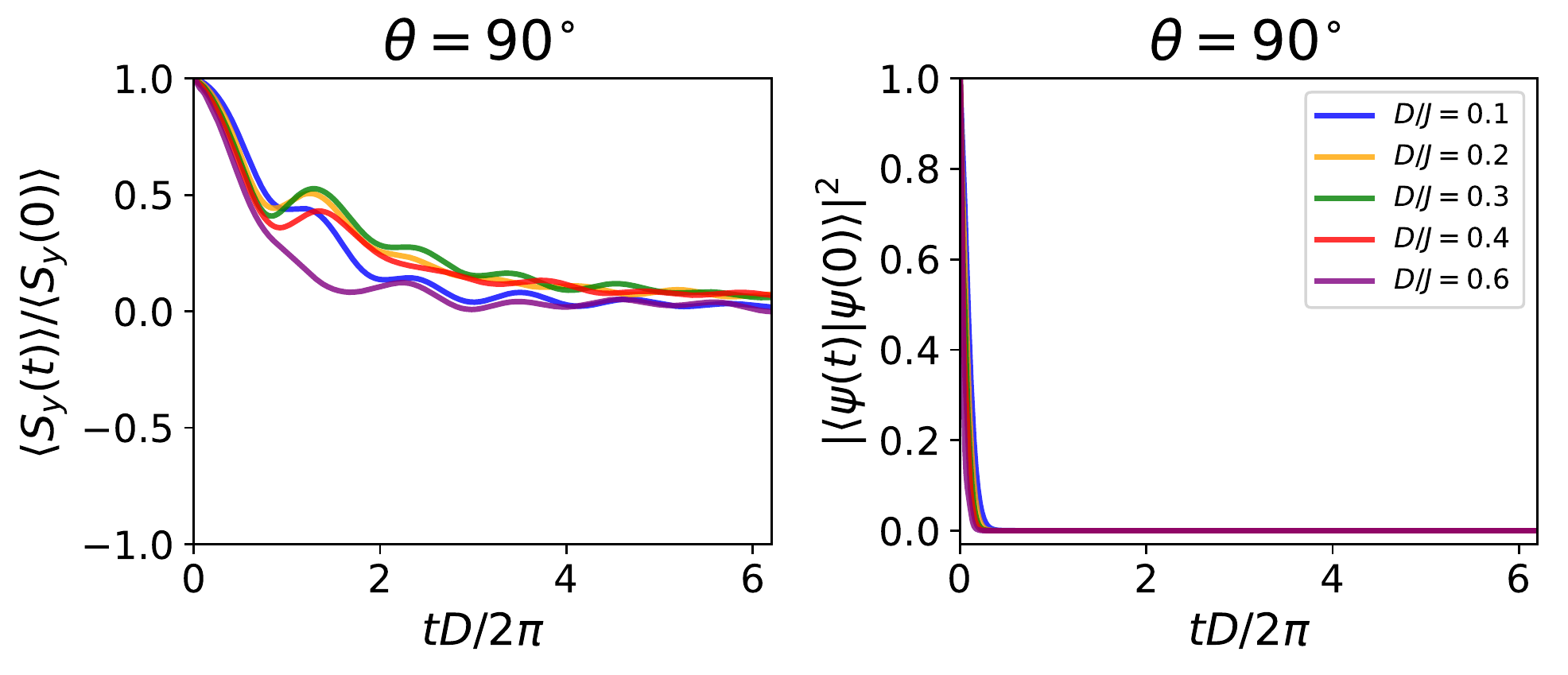}}
\caption{%
Time profiles of $\langle S_y(t) \rangle/ \langle S_y(0) \rangle $ and the Loschmidt echo (revival fidelity)
for various representative $\theta$, corresponding to different starting states $|\psi (\theta,t=0) \rangle$,
calculated using TEBD on spin chains with open boundary conditions.
A maximum bond dimension of $m=100$ was used for $\theta \leq 20^{\circ}$ ($L=200$), and $m=50$ ($L=100$)
was used for the rest of the calculations.
(Times beyond which the bond dimension reaches the maximum value $m$ with truncation cutoff of $\epsilon = 10^{-8}$ are not shown to rule out 
any truncation errors from the growth of entanglement).
In each panel representative small $D/J$ and large $D/J$ values at fixed $\theta$ are shown.
Thermalization is slow (or virtually absent on the time scale of the plot) for small $\theta$ and small $D/J$,
but becomes rapid when either parameter is made large.
}
\label{fig:const_theta}
\end{figure*}

\para{}
We now consider the implications of our findings on recent ultracold atomic experiments performed with the same spin Hamiltonian as in Eq.~\eqref{eq:spin1ham}
(the sign convention of $D$ in Ref.~\onlinecite{chung2021tunable} is the opposite of what we have considered here and elsewhere~\cite{chauhan2019probing}).
The authors of Ref.~\onlinecite{chung2021tunable} implemented this Hamiltonian using a Mott insulator of doubly occupied sites and demonstrated the dynamical properties associated with the presence of single-ion anisotropy.
In this setup, two hyperfine states of $^{87}$Rb, denoted by $\ket{a}$ and $\ket{b}$, are mapped to $S=1$ degrees of freedom via
$S_i^{z} = \frac{1}{2}(a_i^{\dagger}a_i - b_i^{\dagger}b_i)$, $S^{+}_i = a^{\dagger}_i b_i$,
$S^{-}_i = b^{\dagger}_i a_i$,
where $a_i$ and $b_i$ are boson annihilation operators at site $i$ for $\ket{a}$ and $\ket{b}$, respectively, with the constraint that $a^{\dagger}_i a_i + b^{\dagger}_i b_i = 2$.
This mapping between hyperfine states and $S = 1$ degrees of freedom is depicted in Fig.~\ref{fig:single-magnon-dispersion}(c).
We henceforth refer to this setup and the associated experiment as the ``MIT experiment.''

\para{}
The MIT experiment studied spin dynamics by first preparing the state of all atoms as an equal superposition of $\ket{a}$ and $\ket{b}$, using a combination of microwave pulses,
\begin{subequations}
\begin{eqnarray}
	|\psi \rangle &=&
  \bigotimes_{i=1}^{L} \Big( \frac{|a\rangle - i |b\rangle} {\sqrt{2}} \Big)_{i,\text{atom1}} \!\! \otimes \Big( \frac{|a\rangle - i |b\rangle} {\sqrt{2}} \Big)_{i,\text{atom2}} \qquad \\
			  &=&  \bigotimes_{i=1}^{L} \Big( \frac{1}{2} |1\rangle  - i \sqrt{2} |0\rangle - |-1\rangle \Big).
\end{eqnarray}
\end{subequations}
When written out in terms of spin degrees of freedom, this wavefunction is a superposition of multiple magnon sectors with the most dominant contribution coming from the Hilbert space that corresponds to $S_z=0$.
This initial state was allowed to time-evolve and the operator $A = 2-3\frac{1}{L} \sum_{i=1}^{L} \langle (S^{z}_i)^2 \rangle$ was measured as a function of time.
(The material equivalent of the above experiment will require measurements of oscillation and thermalization time scales $\approx 2\pi \hbar/D$ of the order of 10 picoseconds.)

\para{}
Motivated by the MIT experiment, we propose a modification with the objective of demonstrating the importance of magnon-magnon interactions and magnon density on spin dynamics and thermalization.
We prepare an initial state which corresponds to spins rotated about the $x$-axis by angle $\theta$ with respect to the $z$-axis, i.e., with direction vector $(0,\sin \theta, \cos \theta)$.
(In our notation, the angle realized in the MIT experiment is $\theta = -90^{\circ}$.)
The starting ket is given by
\begin{align}
  |\psi (\theta, t=0) \rangle
    = 
      \bigotimes_{i=1}^{L}  \Big( \frac{\cos \theta + 1}{2} |1 \rangle  + \frac{i \sin \theta}{\sqrt{2}} |0 \rangle + \frac{\cos \theta - 1}{2} | -1 \rangle \Big).
\label{eq:initial_state}
\end{align}
Rotation by an arbitrary angle $\theta$ (which can be controlled by applying the microwave pulse for a shorter duration) has the effect of introducing a tunable finite density of magnons.
(We will drop the $\theta$ label in $|\psi (\theta,t) \rangle$ from here on for brevity.)
This initial product state is a linear combination of states with definite magnon number $n$,
\begin{align}
	|\psi (t=0) \rangle =  \sum_{n} P_{n} |\psi\rangle = \sum_n c_n |n \rangle
\end{align}
where $P_{n}$ is the operator which projects the wavefunction to the $n$-magnon sector, and $|n \rangle$ is the $n$-magnon wavefunction whose amplitude is given by $c_n=\langle n | \psi \rangle$.

\para{}
What should one expect to observe in the above setup given the framework developed in the previous sections?
If the magnons were truly noninteracting, the energy spacings between the $n$- and $n+1$-magnon sectors would be exactly $D$.
This has a direct measurable consequence in the time domain.
The measurement of the $S_y$ operator as a function of time
$\langle S_y (t) \rangle \equiv \langle \psi(t) | S_y | \psi (t) \rangle$
would yield the characteristic frequency $E_{n+1} - E_{n} = D$, i.e., the oscillation time scale $T = 2 \pi /D$.
However, magnon-magnon interactions renormalize this energy difference and hence corresponding time period.
At low magnon density (small $\theta$) the magnons are essentially noninteracting.
At higher magnon density, this energy difference decreases due to magnon attraction, and thus a larger time scale of oscillation is expected.

\para{}
While oscillations do dominate the short time behavior, signatures of thermalization are to be expected at long times.
This manifests itself in multiple metrics, for example, at large $t$, $\langle S_y (t) \rangle \rightarrow 0$ and the Loschmidt echo (revival fidelity)
$\left| \langle \psi(0) | \psi (t) \rangle \right|^2 \rightarrow 0$.
(We note that the results for $\theta<0$ are directly related to the case of $\theta>0$. For example, $\langle S_y (t)\rangle$
differs by an overall minus sign for $\theta \rightarrow -\theta$, and thus we discuss only the case of $\theta >0$.)

\para{}
To go beyond the qualitative arguments presented above, we perform matrix product state-based second order TEBD calculations, preparing $|\psi (0) \rangle$ as in Eq.~\eqref{eq:initial_state}. A maximum bond dimension of $m=100$ and a time step of $t=0.02$ (in units of $J=1$) were employed.
We rescale the time axis to be in units of $tD/2\pi$.
In these units, the maxima of the Loschmidt echo and $\langle S_y(t) \rangle$ must occur at every integer for perfectly non interacting magnons.

\para{}
Figure~\ref{fig:const_theta} shows our results for various representative values of $\theta$ and confirms many of our qualitative expectations.
At large $\theta$, i.e., high average density of magnons, (see $\theta = 30^{\circ}$ and $\theta = 90^{\circ}$), only a few (or no) coherent oscillations are observed, and thermalization is rapid.
At small $\theta$ (i.e., low average magnon density), on the other hand, the oscillation time is nearly $2 \pi/D$--a more refined renormalized estimate can be obtained with an approximation we will discuss shortly.
Our investigations suggest that there is a possibility of a prethermal phase~\cite{Langen_2016, Mori_2018, Mallayya_2019} for small $\theta$-dependent $D/J$
(for example,
$D/J \lesssim 0.33$ for $\theta = 10^{\circ}$,
$D/J \lesssim 0.17$ for $\theta = 15^{\circ}$,
and $D/J \lesssim 0.11$ for $\theta = 20^{\circ}$.).
Such long thermalization time scales occur in systems that are close to integrability or have scar-like states in the spectrum.
(See, for example, Refs.~\onlinecite{turner_2018_np, Moudgalya_scar, choi_2019_prl, schecter_2019_prl, lee_2020_prb, lee_prb_fragmentation,
mcclarty_scars, Schoutens_2020, Surace, chertkov2021motif}.)

\para{}
A qualitative explanation of this effect is as follows.
If the prepared initial state has finite overlaps with eigenstates which form a tower of states (states uniformly spaced in energy), it will result in perfectly coherent oscillations in several time-dependent observables.
This arises due to the precession of a superspin of length $L$ (for $S=1$), whose $2L+1$ $S_z$ projected states
(appropriately normalized) are $P_n | \psi \rangle$ for $n=0,1, \ldots, 2L$.
For $D=0$, all $P_n | \psi \rangle$ are exactly degenerate as a consequence of SU(2) symmetry.
For $D/J$ small but nonzero, these states (which, strictly speaking, do not remain exact eigenstates)
have a spacing which is approximately (but not exactly) $D$.
Most importantly, these energy spacings are nonuniform, which, in turn, leads to thermalization in the large time limit, the smaller the nonuniformity the longer the thermalization scale.

\begin{figure*}[htpb]
\subfigure[]{\includegraphics[width=0.33\linewidth]{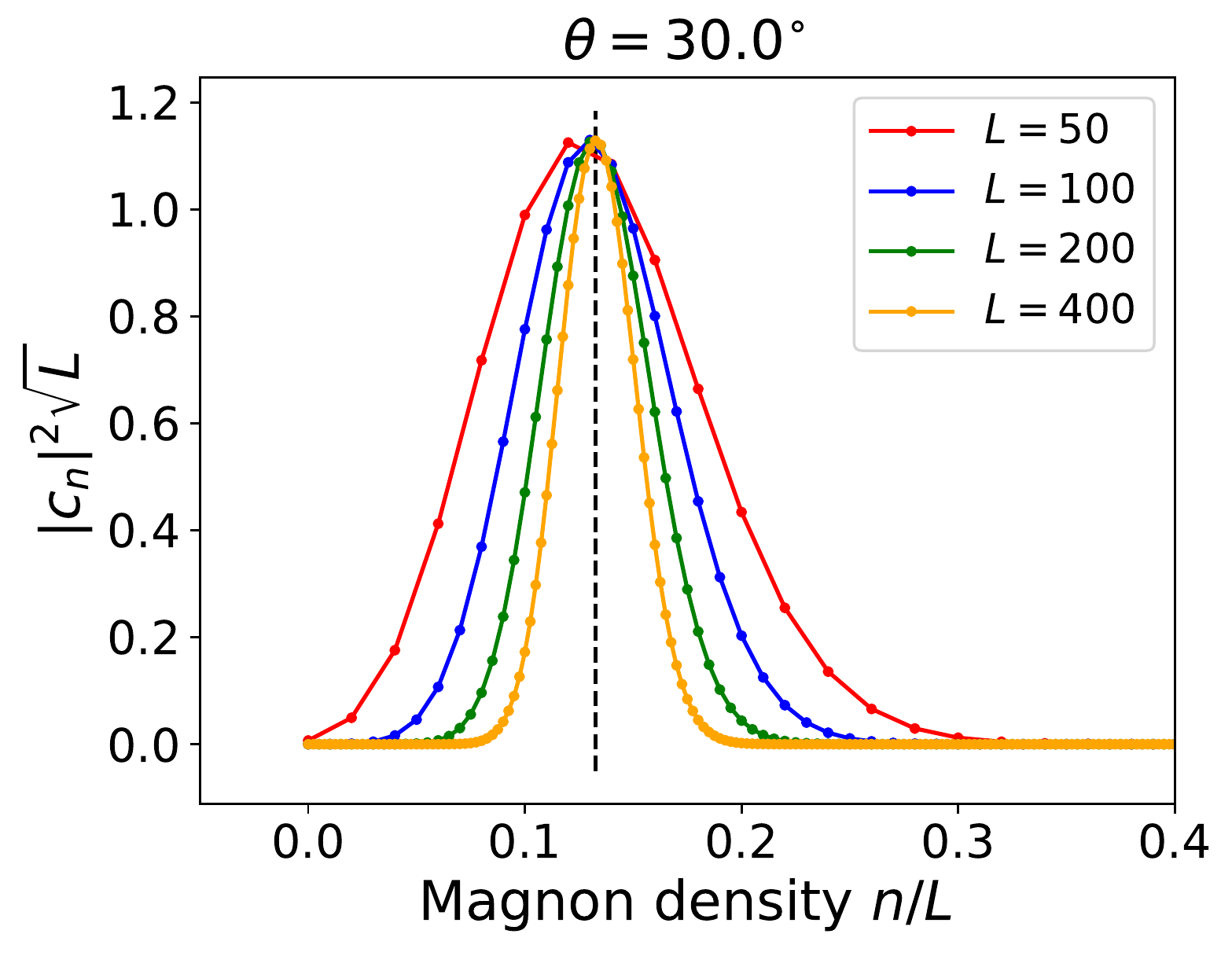}}
\subfigure[]{\includegraphics[width=0.34\linewidth]{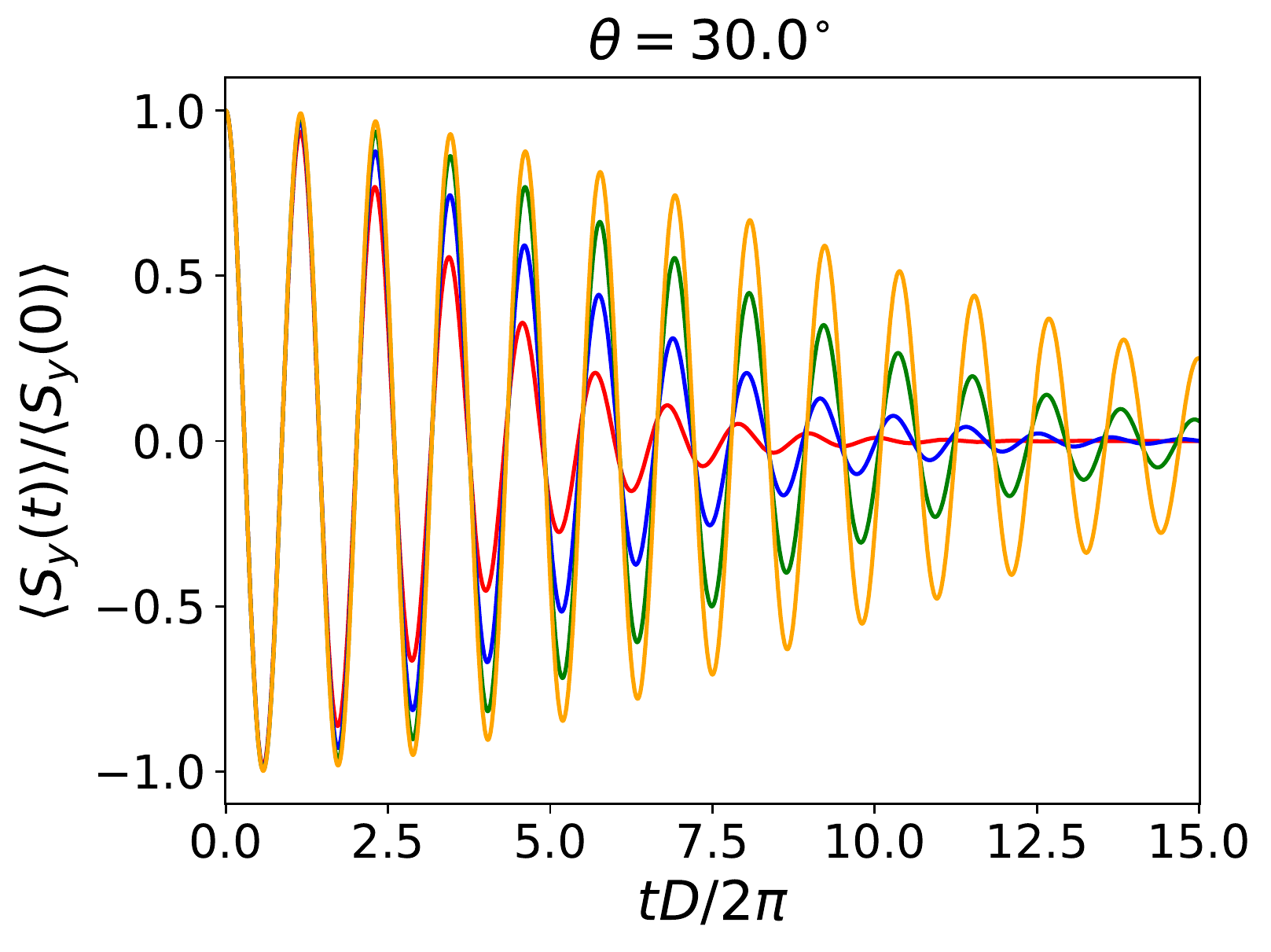}}
\subfigure[]{\includegraphics[width=0.32\linewidth]{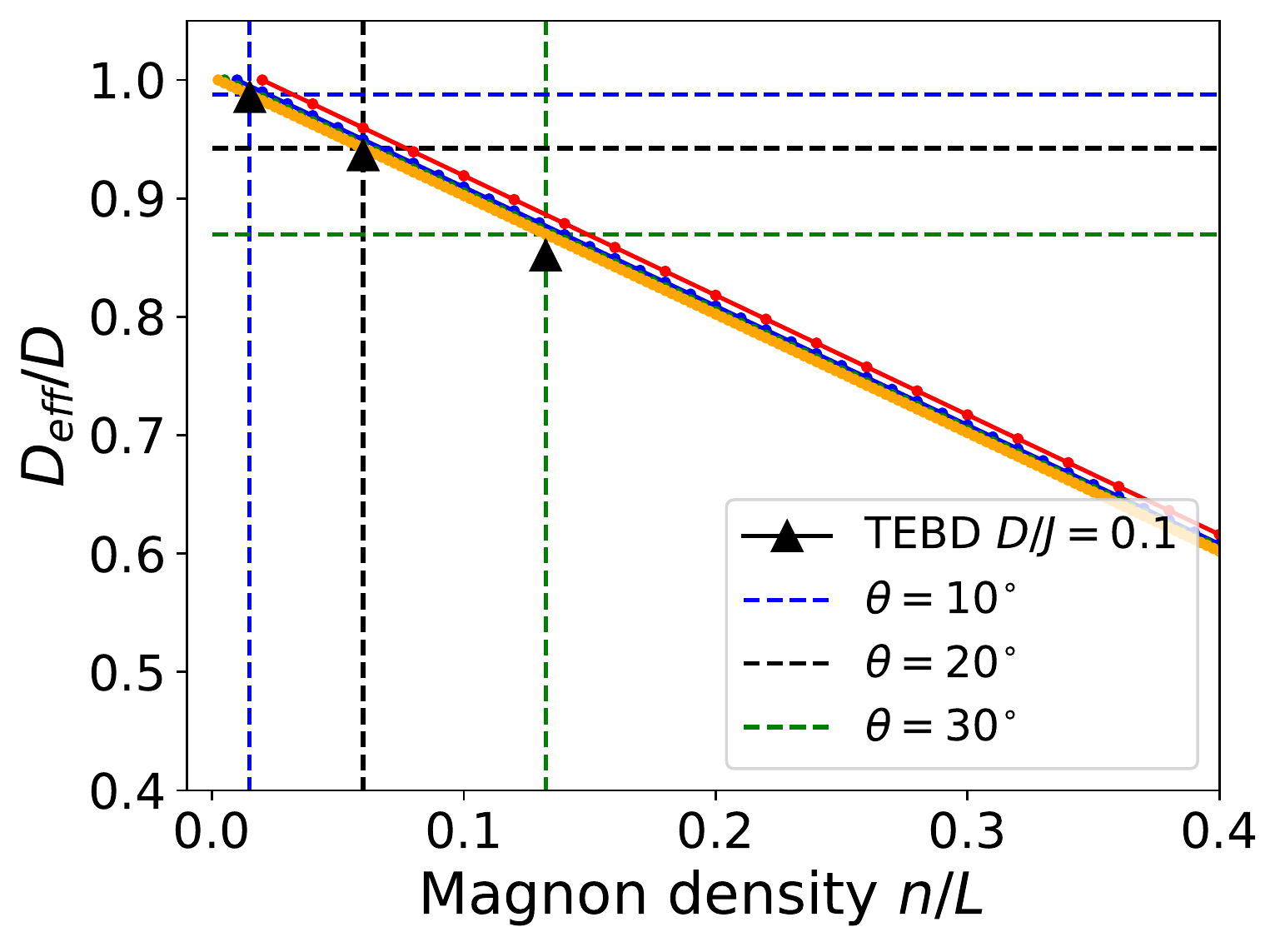}}
\caption{%
Representative results from the approximation in Eq.~\eqref{eq:approx1}.
(a) The amplitude $|c_n|^2 \sqrt{L}$ associated with every magnon sector for $\theta=30^{\circ}$
for representative $L$.
(b) $\langle S_y(t)\rangle/\langle S_y(0) \rangle$ for $\theta=30^{\circ}$.
(c) $D_{\text{eff}}/D$ as a function of magnon density;
determination of $\bar{D}_{\text{eff}}/D$ using the average magnon density $\bar{n}/L$ for the case of $\theta=10^{\circ},20^{\circ},30^{\circ}$
is represented by the dashed lines. Comparison with the $\bar{D}eff/D$ extracted from the period of the first      
oscillation seen  in TEBD (for D/J=0.1) is also shown.
}
\label{fig:approximation}
\end{figure*}

\para{}
We demonstrate these arguments with an approximation which is quantitatively accurate for short time, especially for small $D/J$ and small $\theta$.
In this limit, the eigenstates can be considered essentially unchanged from those for the $D=0$ model.
Thus, we have
\begin{align}
	\exp (-i H t) P_n | \psi \rangle \approx \exp (- i \tilde{E}_n t) P_n | \psi \rangle
	\label{eq:approx1}
\end{align}
where $\tilde{E}_n \equiv \langle \psi | P_n H P_n | \psi \rangle / \langle \psi| P_n P_n | \psi \rangle$ is the energy of the state $P_n |\psi \rangle$.
The computation of arbitrary operator expectation values within this approximation is straightforward:
For example, $\langle S_y (t) \rangle$ is
\begin{subequations}
\begin{align}
  \langle S_y (t) \rangle
    &= \sum_{n,m} \langle \psi | P_n e^{i H t} S_y e^{-iH t} P_m | \psi \rangle  \\
    &\approx \sum_{n,m} c_n^{*} c_m e^{i (\tilde{E}_n - \tilde{E}_m) t} \langle n | S_y | m \rangle
\end{align}
\end{subequations}
(Similar computations can be carried out for other operators using the algebra of coherent states, see for example~\cite{Changlani_PRL_2018,lee_2020_prb, Pal_PRB_2021}.)

\para{}
The only nonzero contributions to $\langle S_y(t) \rangle$ are from $m=n\pm1$.
Note that $\langle n | \sum_{\langle i,j \rangle} \bfS_i \cdot \bfS_j | n \rangle$ for an $L$-site periodic chain equals $L$,
i.e., it is independent of $n$, which follows from the fact that the normalized
$|n\rangle \propto P_n |\psi \rangle$ are the different $S_z$ projections of a superspin of length $L$.
Hence the energy difference $\tilde{E}_{n} - \tilde{E}_{n\pm1}$ arises purely from the on-site anisotropy term and does not depend on $J$.
Hence it is convenient to define
\begin{subequations}
\begin{align}
  D_{\text{eff}}^{(n)}
    &\equiv \tilde{E}_{n} - \tilde{E}_{n-1}\\
    &=
      - D  \sum_i  \Big(
        \langle n |(S_{i}^{z})^2 | n \rangle
        - \langle n\mathord{-}1 | (S_{i}^{z})^2 | n\mathord{-}1 \rangle
      \Big).
\end{align}
\end{subequations}
An exact computation yields,
\begin{align}
  \Big\langle n \Big| \sum_i (S_{i}^{z})^2 \Big| n \Big\rangle
  &=
    L - n
    + 2\frac{\sum_{m=0}^{[n/2]}
        \frac{m}{2^{2m}} \binom{L}{L-n+m,n-2m,m}
      }{
        \sum_{m=0}^{[n/2]} \frac{1}{2^{2m}} \binom{L}{L-n+m,n-2m,m}
      }
\end{align}
where $\binom{L}{i,j,k}= \frac{L!}{i!j!k!}$ for $i+j+k=L$ is the trinomial function.
Note that $\left\langle n \middle| \sum_i (S_{i}^{z})^2 \middle| n \right\rangle$ and hence the plot for $D_{\text{eff}}^{(n)}/D$ versus magnon density $n/L$ is independent of $\theta$.
$\theta$ has the effect of selecting the average magnon density and hence the value of $D_{\text{eff}}^{(n)}$ which controls the time period of the oscillations.

\para{}
Figure~\ref{fig:approximation} shows representative results for periodic chains within the framework of the approximation.
For $\theta=30^{\circ}$, the amplitude $|c_n|^2 \sqrt{L}$ associated with every magnon sector for representative $L$ is shown as a function of magnon density.
(The profile is expected to approach Gaussian, and hence the factor $\sqrt{L}$ is introduced when comparing different system sizes.)
Since the system sizes are finite, nonzero contributions arise from a range of magnon densities.
In the thermodynamic limit, however, the only nonzero contribution will be from $\bar{n}/L$, the average magnon density.
$\theta$ thus controls the average magnon density in the wavefunction $|\psi \rangle$.

\para{}
$\bar{n}/L$ in turn determines $\bar{D}_{\text{eff}}$, the $D_{\text{eff}}^{(n)}$ with the largest contribution.
The right panel of Fig.~\ref{fig:approximation} shows this connection with the help of dashed lines for the case of $\theta=10^{\circ},20^{\circ},30^{\circ}$.
The value of $\bar{D}_{\text{eff}}/D$ decreases on increasing $\theta$ (and hence magnon density).
Since this $\bar{D}_{\text{eff}}$ sets the time period of oscillations $T \approx 2\pi/\bar{D}_{\text{eff}}$, we are able to infer its value from the TEBD calculations.
We compare the TEBD result for the time of the first oscillation for $D/J=0.1$ with the approximate result and confirm that the discrepancy is less than a percent.
However, the approximation does not capture higher order effects in $D$ and $\theta$ and long time behavior, for example, in the (exact) TEBD calculations oscillations are not seen for large $\theta \gtrsim 30^{\circ}$. (See Fig.~\ref{fig:const_theta}.)

\para{}
The central panel shows $\langle S_y(t) \rangle/ \langle S_y(0) \rangle$ calculated within our approximation.
For short time, the finite size effects are essentially negligible.
At longer times, $\langle S_y(t) \rangle/ \langle S_y(0) \rangle$ appears to decay, but this is a finite size effect.
All nonzero contributions to $\langle S_y(t)\rangle$ originate from a single $D_{\text{eff}}$ in the thermodynamic limit;
for any finite size system there is always a spread of contributing energy scales as discussed previously in reference to $|c_n|^2 \sqrt{L}$.
Hence there is no thermalization in this approximation for the infinite chain limit.

\para{}
Since the short and long time behavior depend on both $\theta$ and $D/J$, we also plot our TEBD data for various representative values of $D/J$ and analyze their dependence on $\theta$ in Appendix~\ref{sec:dynamics1}.
We find that the time period increases with $\theta$ (magnon density) consistent with the reduction of $\bar{D}_{\text{eff}}$ the energy spacing between magnon sectors.

\section{Conclusion}
\label{sec:conclusion}

\para{}
In summary, we have studied the energetics, dynamics and thermalization of multiple interacting magnons in a $S=1$ chain with ferromagnetic Heisenberg interactions and easy-axis anisotropy using both analytic and numerical methods.
The model and its analyses presented here are of direct relevance to both real materials~\cite{chauhan2019probing} and cold-atom setups~\cite{chung2021tunable}, where different aspects of the dynamics have been recently investigated.

\para{}
Building on previous literature~\cite{tonegawa_1970_ptps, majumdar1970application},
we established that the easy-axis anisotropy ($D>0$ in this paper) serves as a source of attractive magnon-magnon interactions which leads to the formation of magnon clouds, whose characteristics we explored.
Many of the properties of these clouds are captured by a simple pair Jastrow function that shows good agreement with numerical (almost exact) DMRG results.
For a small number of magnons, the cloud does not significantly alter its size (spatial extent) on the introduction of additional magnons.
The energy cost for having additional magnons decreases from $D$ towards zero, and once the magnon cloud is saturated entirely, 
the formation of Ising-type bound states becomes important, which manifests itself as an even-odd magnon number effect in the energy cost for adding a magnon.

\para{}
Importantly, the lessons learnt from the energetics of the few magnon problem were used to clarify the origin of the temperature-dependent frequency shift observed in THz dynamical susceptibility measurements~\cite{chauhan2019probing}. 
The dynamical Kubo formula involving matrix elements and energy scales was analyzed, and the energy scales effective at high temperatures were identified.
The reduction of the effective value of $D$ with magnon density demonstrated the importance of magnon-magnon interactions on nonequilibrium dynamics in the time domain 
in quench experiments that initialized the system in a superposition of multimagnon states. An attractive feature of the cold-atom setup that realizes this protocol is that both the magnitude and sign of the single-ion anisotropy can be tuned~\cite{chung2021tunable}, and the average density of magnons can also be potentially controlled.
With the help of matrix product state-based TEBD calculations, we studied the time evolution of the $S_y$ expectation value
(which is sensitive to gaps between consecutive magnon sectors, and hence magnon-magnon interactions) and the Loschmidt echo in order to develop an understanding of revivals and thermalization in this model.
Many of the observed behaviors are akin to those noted in the context of quantum scars; we showed how a simplified superspin picture explains our results for small $D/J$.

\para{}
It would be interesting to realize the possibility of experimental measurements that verify the picture we have developed here.
On this front, time dependent THz measurements offer a potentially exciting route for studying the thermalization of spin chains.
It would also be valuable to model the evolution of magnon clouds created in a small portion of the lattice, for example, by applying a microwave pulse only on a section of the optical lattice.
Further development of a quantitative analytic framework for understanding the $D/J$ dependence of the prethermalization and thermalization time scales observed in our TEBD calculations should be relevant for a wide variety of other realistic systems where athermal (or nearly athermal) states exist, and which show unusually slow or glassy dynamics.
Finally, the study of bound states in higher spin chains may also be of interest, as has been recently studied in Ref.~\cite{wu2021fewmagnon}.

\begin{acknowledgments}
We thank F. Mahmood, P. Chauhan, P. Armitage and W. Chung for insightful discussions about several aspects of the JHU and MIT experiments.
We thank Florida State University and the National High Magnetic Field Laboratory for start up funds. The National High Magnetic Field
Laboratory is supported by the National Science Foundation through NSF/DMR-1644779 and the state of Florida.
H.J.C. was also supported by NSF CAREER grant DMR-2046570.
The DMRG and TEBD calculations were performed using the ITensor C++ library (version 2.1.1)~\cite{itensor}.
We also thank the Research Computing Cluster (RCC) and Planck cluster at Florida State University for computing resources.
\end{acknowledgments}

\appendix

\section{\texorpdfstring{$t$}{t}-matrix approximation for two-magnon bound state}
\label{sec:t-matrix}

\subsection{Basic definitions}

\para{}
Consider a bosonic Hamiltonian
\begin{align}
  H = H[a^{\dagger}, a].
\end{align}
The quadratic part of the Hamiltonian, with translation symmetry, can be written as
\begin{align}
  H_0 = \sum_{q} \epsilon_{q} a_{q}^{\dagger} a_{q}
\end{align}
where $a_{q}^{\dagger} = \frac{1}{\sqrt{L}} \sum_{x} a_{x}^{\dagger} e^{i q x}$.
Using the imaginary time evolved $a_{q}$ and $a_{q}^{\dagger}$
\begin{align}
  a_{q}(\tau) &= e^{\tau H} a_{q} e^{- \tau H}, &
  a_{q}^{\dagger}(\tau) &= e^{\tau H} a_{q}^{\dagger} e^{- \tau H},
\end{align}
the imaginary time-ordered bosonic Matsubara Green function can be defined as
\begin{align}
  \calD (q, \tau) &\equiv - \langle \calT_{\tau} a_{q}(\tau) a_{q}^{\dagger}(0) \rangle
\end{align}
where the bracket indicates the thermal expectation value
\begin{align}
  \langle O \rangle &= \frac{\mathrm{Tr} \left( e^{-\beta H} O \right) }{\mathrm{Tr}\; e^{-\beta H} }.
\end{align}
The imaginary time Green's function can be written in terms of Matsubara frequency $i\omega_{n}$
\begin{align}
  \calD(q, i\omega_{n}) &= \frac{1}{2} \int_{-\beta}^{\beta} \! \rmd \tau \; e^{i \omega_n \tau} \calD(q, \tau)
\end{align}
The noninteracting Green function in terms of Matsubara frequency writes
\begin{align}
  \calD^{(0)} (q, i\omega_{n}) = \frac{1}{i\omega_{n} - \epsilon_{q}}.
\end{align}

\para{}
The interaction term in general can be written as
\begin{subequations}
\begin{align}
  \hat{V}
    &=
       \sum_{ \{ x_{i} \} }
        V{}^{x_1 x_2}_{x_3 x_4}
        a_{x_{1}}^{\dagger} a_{x_{2}}^{\dagger} a_{x_{4}} a_{x_{3}}
    \\
    &=
      \frac{1}{L}  \sum_{ \{ q_{i} \} }
        V{}^{q_{1} q_{2}}_{q_{3} q_{4}}
        a_{q_{1}}^{\dagger} a_{q_{2}}^{\dagger} a_{q_{4}} a_{q_{3}},
\end{align}
\end{subequations}
where $L$ is the number of sites of the system, and
\begin{align}
  V{}^{q_{1} q_{2}}_{q_{3} q_{4}}
    &=
      \frac{1}{L}
      \sum_{ \{ x_{i} \} }
      V{}^{x_1 x_2}_{x_3 x_4}
      e^{- i ( q_{1} x_{1} + q_{2} x_{2} - q_{3} x_{3} - q_{4} x_{4} ) }
\end{align}
Note that the matrix element of the interaction in terms of the positions $V{}^{x_1 x_2}_{x_3 x_4}$,
and in terms of momenta $V^{q_1 q_2}_{q_3 q_4}$ both are of order $O(1)$ with respect to the number of sites.
With translation symmetry, the interaction term conserves total momentum:
\begin{align}
  V{}^{q_1 q_2}_{q_3 q_4}
    &= \delta_{q_1 + q_2 = q_3 + q_4} V{}^{q_1 q_2}_{q_3 q_4}
\end{align}

\subsection{Feynman rules and \texorpdfstring{$t$}{t}-matrix approximation}

\para{}
The Feynman rules for interacting bosons $a$ and $a^{\dagger}$ are
\begin{align}
  \begin{tikzpicture}[baseline=(a)]
    \begin{feynman}
      \vertex (a) at (-1, 0) {};
      \vertex (b) at ( 1, 0) {};
      \diagram* {
        (a) -- [fermion, edge label={$q$}] (b),
      };
    \end{feynman}
  \end{tikzpicture}
&\equiv - \calD^{(0)} ( \bfq, i\omega_{n} ) = - \frac{1}{i\omega_{n} - \epsilon_{\bfq}}
\\
  \begin{tikzpicture}[baseline=(m)]
    \begin{feynman}
      \vertex[crossed dot] (m) at ( 0, 0) {};
      \vertex (a) at (-0.75,-0.75) {$q_3$};
      \vertex (b) at ( 0.75,-0.75) {$q_2$};
      \vertex (c) at (-0.75, 0.75) {$q_4$};
      \vertex (d) at ( 0.75, 0.75) {$q_1$};
      \diagram* {
        (a) -- [fermion] (m) -- [fermion] (b),
        (c) -- [fermion] (m) -- [fermion] (d),
      };
    \end{feynman}
  \end{tikzpicture}
  &\equiv
    - \makebox{$V$}^{q_{1} q_{2}}_{q_{3} q_{4}}
\end{align}
with energy momentum conservation
\begin{align}
  \int_{q} \equiv
  \frac{1}{L} \sum_{\bfq} \frac{1}{\beta} \sum_{i \omega_{n}}
    \xrightarrow{T \rightarrow 0, L \rightarrow \infty}
    \int \! \frac{\rmd^{d}q}{(2\pi)^d} \frac{\rmd\omega}{(2\pi)}
\end{align}

The Feynman diagrams of two magnon susceptibility can be expanded in $V$ as
\begin{align}
\chi_{2}
  &=
  \feynmandiagram[small, inline=(a)] {
    a [dot] -- [fermion, quarter left ] b [dot],
    a -- [fermion, quarter right] b,
  };^{\checkmark}
  +
  \feynmandiagram[small, inline=(a)] {
    a [dot]
      -- [fermion, quarter left ] m [crossed dot]
      -- [fermion, quarter left ] b [dot],
    a -- [fermion, quarter right] m
      -- [fermion, quarter right ] b,
  };^{\checkmark}
  \nonumber\\
  &+
  \feynmandiagram[small, inline=(a)] {
    a [dot]
      -- [fermion, quarter left ] m1 [crossed dot]
      -- [fermion, quarter left ] m2 [crossed dot]
      -- [fermion, quarter left ] b [dot],
    a -- [fermion, quarter right] m1
      -- [fermion, quarter right] m2
      -- [fermion, quarter right ] b,
  };^{\checkmark}
  +
    \begin{tikzpicture}[baseline=(a)]
      \begin{feynman}[small]
        \vertex[crossed dot] (m1) at ( 0, 0.3) {};
        \vertex[crossed dot] (m2) at ( 0, -0.3) {};
        \vertex[dot] (a) at (-0.6, 0) {};
        \vertex[dot] (b) at ( 0.6, 0) {};
        \diagram* {
          (a) -- [fermion, quarter left] (m1) -- [fermion, quarter left] (b),
          (a) -- [fermion, quarter right] (m2) -- [fermion, quarter right] (b),
          (m1) -- [fermion, quarter right] (m2);
          (m2) -- [fermion, quarter right] (m1);
        };
      \end{feynman}
    \end{tikzpicture}
   + \ldots
\end{align}
The summation of the checked diagrams and their higher-order versions is known as the $t$-matrix approximation, which can be expressed concisely as
\begin{align}
  \chi_{2}^{t} &= \chi^{(0)} \cdot (1 + V \cdot \chi^{(0)})^{-1},
\end{align}
where the objects are understood as matrices in terms of momenta, and the dots $(\cdot)$ represent matrix multiplications.
\begin{widetext}
\begin{align}
\chi_{2}^{t} (q)
  &=
  \feynmandiagram[small, inline=(a)] {
    a [dot] -- [fermion, quarter left, edge label={$k\mathord{+}q$}] b [dot],
    a -- [fermion, quarter right, edge label'={$-k$}] b,
  };
  +
  \feynmandiagram[small, inline=(a)] {
    a [dot]
      -- [fermion, quarter left, edge label={$k_1\mathord{+}q$} ] m [crossed dot]
      -- [fermion, quarter left, edge label={$k_2\mathord{+}q$}] b [dot],
    a
      -- [fermion, quarter right, edge label'={$-k_1$}] m
      -- [fermion, quarter right, edge label'={$-k_2$}] b,
  };
  +
  \feynmandiagram[small, inline=(a)] {
    a [dot]
      -- [fermion, quarter left, edge label={$k_1\mathord{+}q$}] m1 [crossed dot]
      -- [fermion, quarter left, edge label={$k_2\mathord{+}q$}] m2 [crossed dot]
      -- [fermion, quarter left, edge label={$k_3\mathord{+}q$}] b [dot],
    a
      -- [fermion, quarter right, edge label'={$-k_1$}] m1
      -- [fermion, quarter right, edge label'={$-k_2$}] m2
      -- [fermion, quarter right, edge label'={$-k_3$}] b,
  };
  +
  \ldots\\
  &=
      \int_{k_1} \! \calD^{(0)}(k_1 + q) \calD^{(0)}(-k_1)
    - \int \!
      \calD^{(0)}(k_1 + q) \calD^{(0)}(-k_1)
      \makebox{$V$}^{k_2+q, -k_2}_{k_1 + q, -k_1}
      \calD^{(0)}(k_2 + q) \calD^{(0)}(-k_2) \nonumber\\
  &\quad + \int \calD^{(0)}\calD^{(0)}V\calD^{(0)}\calD^{(0)}V\calD^{(0)}\calD^{(0)} + \cdots
  \nonumber
\end{align}
\end{widetext}
The interaction matrix element can be expanded in terms of momenta
\begin{align}
  \makebox{$V$}^{k_1 k_2}_{k_3, k_4}
    &= (V_{0} + \calO(k_1, k_2, k_3, k_4)) \delta_{k_1 + k_2 = k_3 + k_4}
  \nonumber\\
    &\approx V_0  \delta_{k_1 + k_2 = k_3 + k_4}.
\end{align}
If the interaction is on-site, then the $t$-matrix two-magnon susceptibility writes
\begin{align}
  \chi_{2}^{t}(q, i\omega)
    &= \frac{\chi^{(0)}(q, i\omega)}{1 + V_0 \chi^{(0)}(q, i\omega)},
\end{align}
where
\begin{subequations}
\begin{align}
  \chi^{(0)}(q, i\omega) &\equiv \int_{k} \calD^{(0)}(k+q) \calD^{(0)}(-k)
  \\
    &=
      \int \! \frac{\rmd^d k}{(2\pi)^d}
      \frac{-1}{i \omega - (\epsilon_{-k} + \epsilon_{k+q})}
\end{align}
\end{subequations}
is the Lindhard susceptibility.

\subsection{Application to the 1D chain}

\para{}
On a one-dimensional chain, if we approximate $\epsilon_k \approx \alpha k^2 + m$,
\begin{align}
  \epsilon_{k+q} + \epsilon_{-k}
    &\approx \alpha (k+q)^2 + m + \alpha k^2 + m \nonumber\\
    &= 2 \alpha \left( k + \frac{q}{2} \right)^2 + \frac{1}{2}\alpha q^2 + 2m
\end{align}
and thus the noninteracting susceptibility becomes
\begin{align}
  \chi^{(0)}(q, i\omega)
    &=
    \frac{1}{2\alpha}
    \frac{1}{2\sqrt{\frac{q^2}{4} + \frac{m}{\alpha} - \frac{i\omega}{2\alpha} }}.
\end{align}
This shows that the condition to have a pole in $\chi^{t}$, which is $V_{0} \chi^{(0)}(\bfq, i\omega)=-1$, can be expressed as
\begin{align}
    \frac{-V_0}{4\alpha} &= \sqrt{\frac{q^2}{4} + \frac{m}{\alpha} - \frac{i \omega}{2\alpha} },
\end{align}
which, after analytic continuation to real frequency $i\omega \rightarrow \omega$, has a solution
\begin{align}
  \omega_{\text{pole}}
    &=  \frac{1}{2} \alpha q^2
       + 2m - \frac{ {V_0}^2 }{8 \alpha}
     = A_{2} q^2 + M_2
\end{align}
The mass of the two-magnon bound state $M_2$ is smaller than the mass of two independent magnons by ${V_0}^2/8\alpha$, which is the binding energy $\Delta$ of a two-magnon bound state when the interaction is attractive.
When the interaction is repulsive, on the other hand, $t$-matrix does not allow a pole.
In terms of the parameters of the original Heisenberg spin-chain Hamiltonian
\begin{align}
  \alpha &\equiv J S, &
  m &\equiv (2 S - 1) D,  & 
  V_0 &= -D,
\end{align}
the binding energy of the magnon is then
\begin{align}
  \Delta = 2 D - M_2 = \frac{D^2}{8J}.
\end{align}

\section{Four-magnon correlator and Jastrow fits}
\label{sec:4magnon}

\begin{figure*}[]
\subfigure[\label{fig:four-magnon-correlator-a}]{\includegraphics[width=0.67\linewidth]{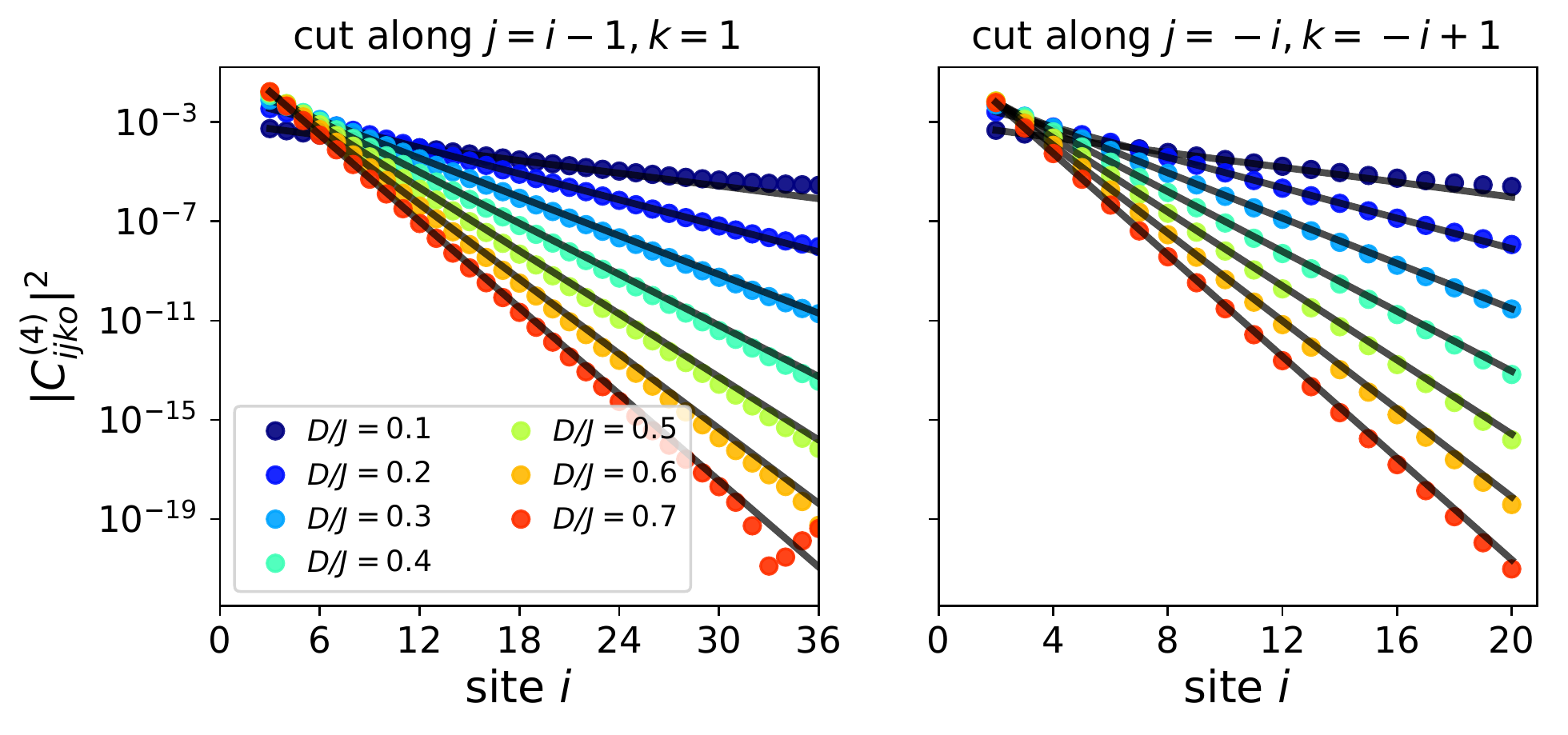}}
\subfigure[\label{fig:four-magnon-correlator-b}]{\includegraphics[width=0.32\linewidth]{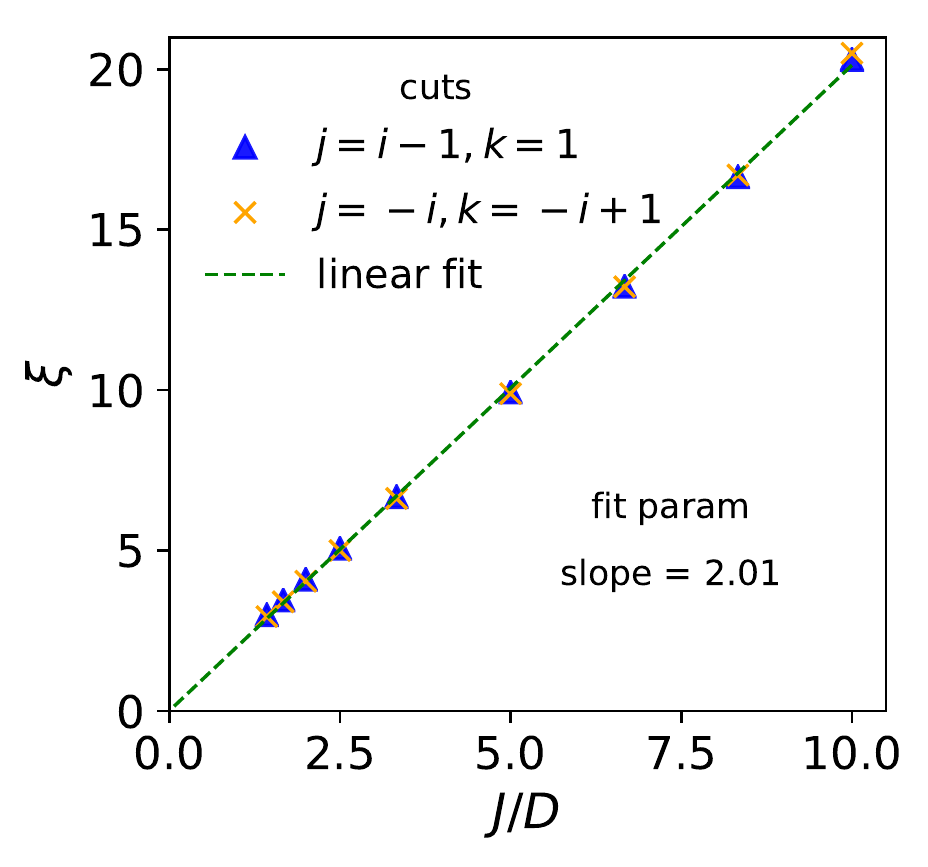}}
\caption{\label{fig:four-magnon-correlator}%
The four-magnon correlator $C_{ijko}^{(4)}$, obtained by fixing one magnon at a reference site $o$,
computed for the ground state of four magnons for representative values of anisotropy parameter $D$, in a periodic chain of length $L=76$.
\subref{fig:four-magnon-correlator-a}
The spatial dependence of the correlator for two one-dimensional cross sections, respectively corresponding to the two magnons at sites $i$ and $j$
with $j=i-1$ and $k=1$ (left panel),
and $j=-i$ and $k=-i+1$ (right panel).
Both cross sections show exponential decay with increasing spatial dependence between the magnons.
Black solid lines are the fits with Jastrow factor $\prod_{a<b} e^{|x_a-x_b|/{\xi}}$.
\subref{fig:four-magnon-correlator-b}
Correlation length $\xi$ vs inverse anisotropy $1/D$ (with $J=1$), obtained from fitting to the Jastrow form.
Just as in the case of three magnons, here too the green dashed line shows that the linear fit is an excellent approximation to the four magnon correlator data.}
\end{figure*}

\para{}
In Sec.~\ref{sec:highermag} we discussed the case of three-magnon wavefunctions and compared the corresponding correlator with the Jastrow theory.
In this Appendix we show the corresponding calculation for the case of four magnons and find good agreement as well.
To visualize the correlator, we fix one magnon at reference site $o$ (the middle of the chain) and consider two representative cross sections, as has been shown in Fig.~\ref{fig:four-magnon-correlator}.

\section{Nonequilibrium dynamics with fixed \texorpdfstring{$D/J$}{D} for different rotation angles}
\label{sec:dynamics1}

\begin{figure*}[]
\subfigure[]{\includegraphics[width=0.49\linewidth]{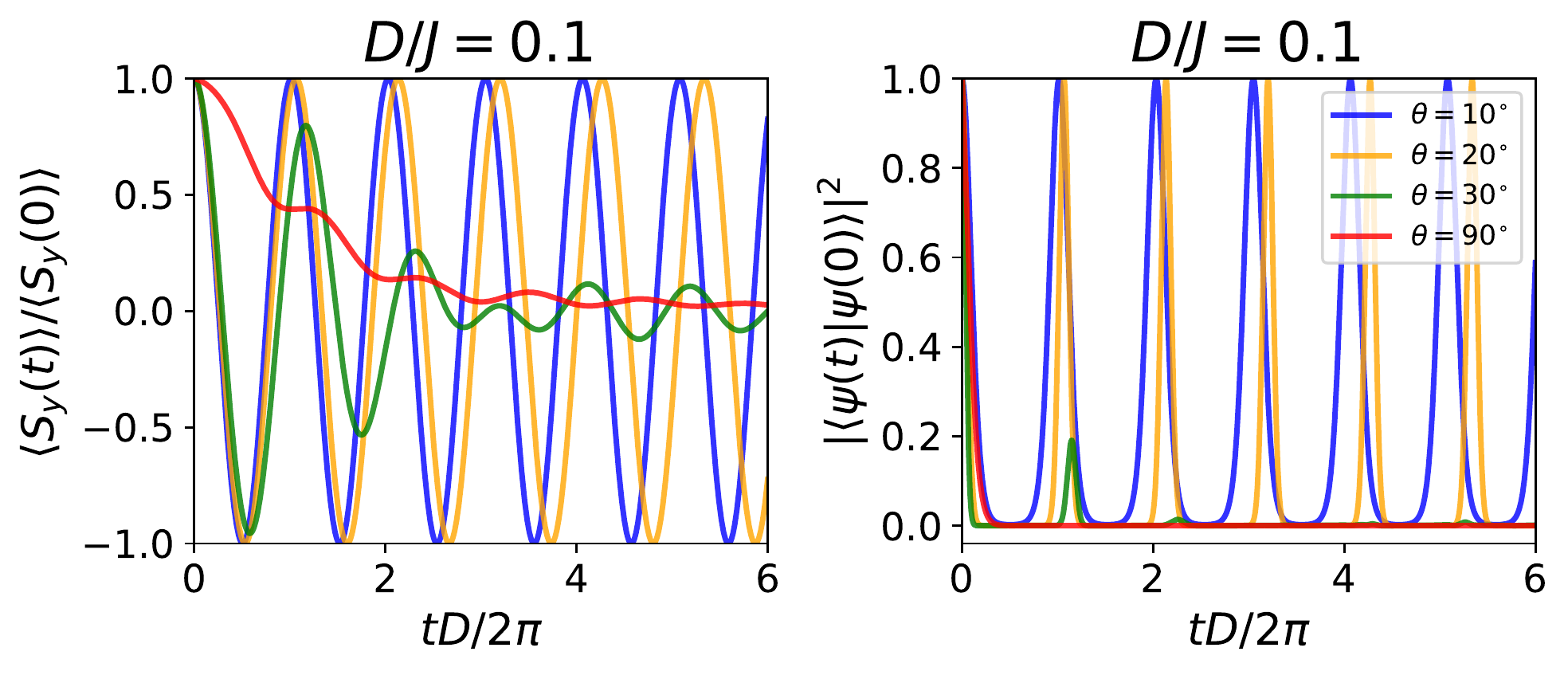}}
\subfigure[]{\includegraphics[width=0.49\linewidth]{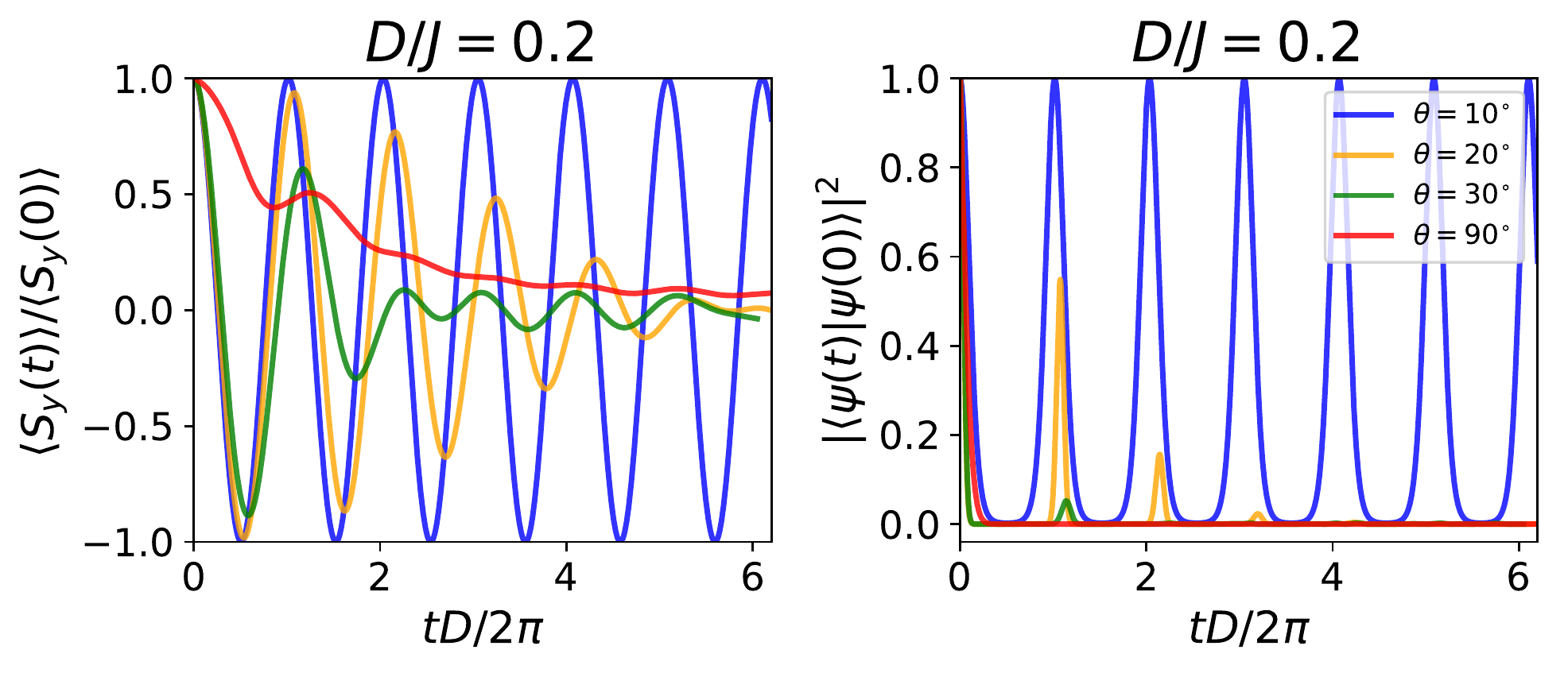}}
\subfigure[]{\includegraphics[width=0.49\linewidth]{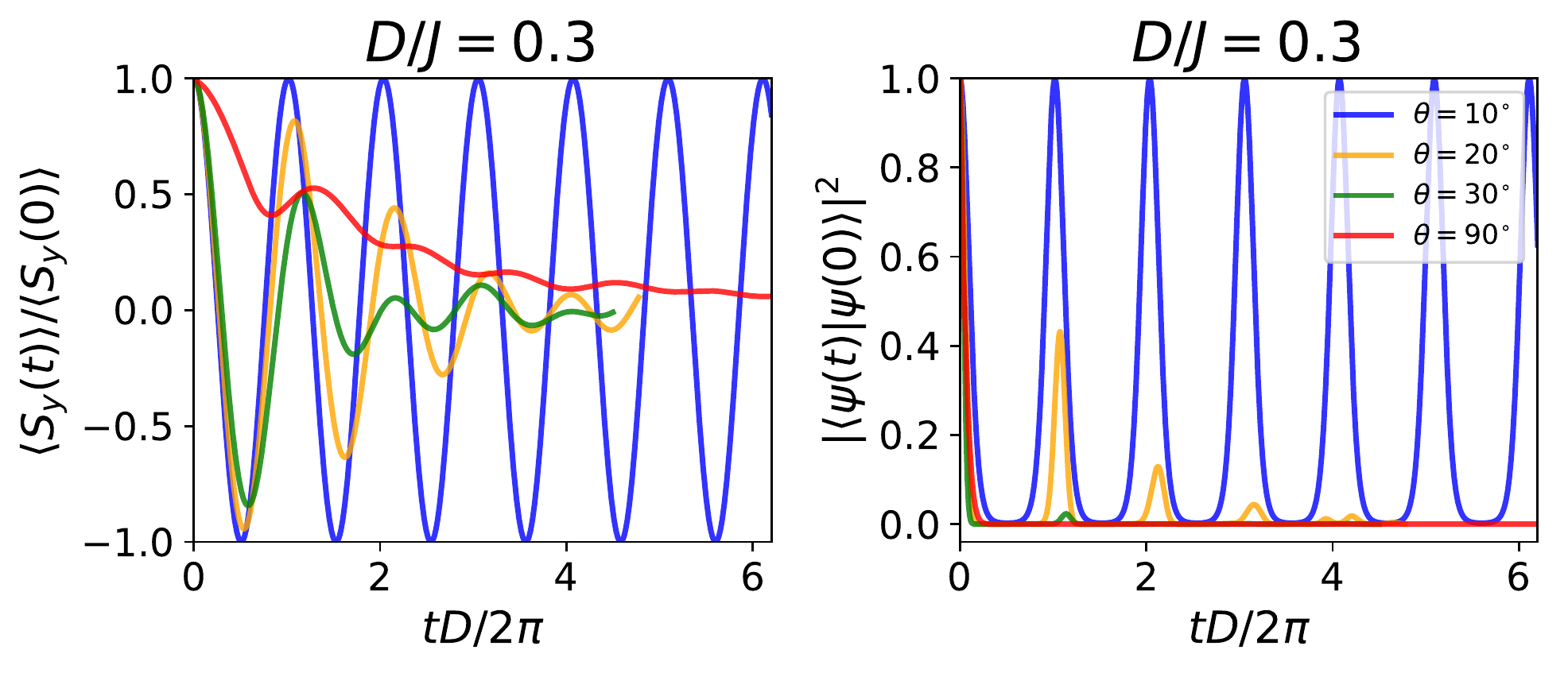}}
\subfigure[]{\includegraphics[width=0.49\linewidth]{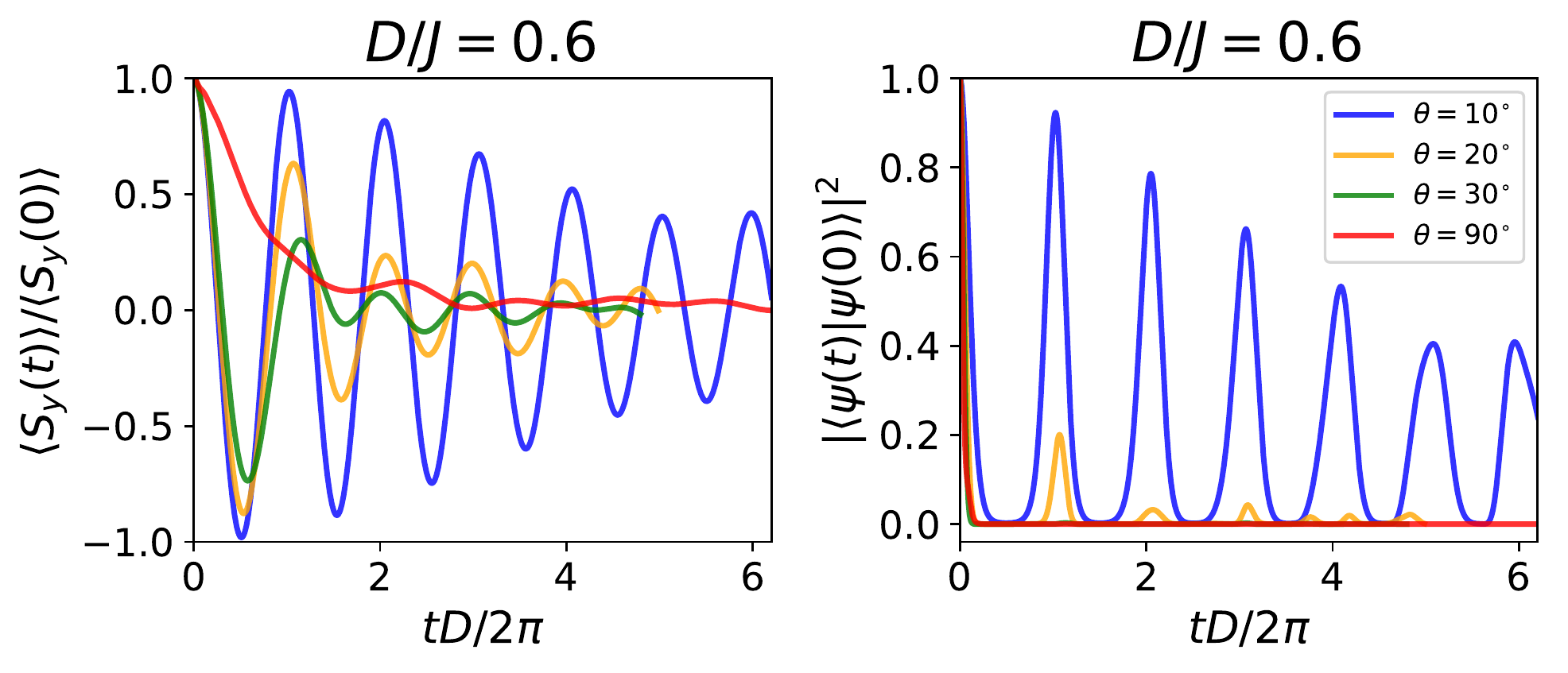}}
\caption{%
Time profiles of $\langle S_y(t) \rangle/ \langle S_y(0) \rangle$ and the Loschmidt echo for various $D/J$.
For each value of $D/J$ representative $\theta$ (which controls the average magnon density) are shown.
A maximum bond dimension of $m=100$ was used for $\theta \leq 20^{\circ}$ ($L=200$), and $m=50$ ($L=100$)
was used for the rest of the calculations. At short times, there is an increase in the effective oscillation time scale on increasing $\theta$
due to a reduction in $D_{\text{eff}}$ as discussed in the text.
}
\label{fig:const_D}
\end{figure*}

\para{}
In Sec.~\ref{sec:scar}, we showed the time profiles of $\langle S_y (t) \rangle$ and the Loschmidt echo
(computed with the TEBD method) for the case of representative $\theta$ and studied their $D/J$ dependence.
In Fig.~\ref{fig:const_D} we plot this (and additional data) for various representative values of $D$ and compare their $\theta$ dependence.
In all cases where an oscillation can be clearly identified, we observe that the time period is larger for larger $\theta$ (i.e., larger magnon density).
This is consistent with the increased role of magnon-magnon attraction, which effectively reduces $\bar{D}_{\text{eff}}$ the spacing between energy levels.

\bibliography{bibliography}

\end{document}